\documentclass[aps,prl,twocolumn,preprintnumbers,amsmath,amssymb,superscriptaddress]{revtex4-2}
\usepackage{amsmath,graphicx}
\usepackage[utf8]{inputenc}
\usepackage[T1]{fontenc}
\usepackage{xcolor}
\usepackage{textcomp}
\usepackage{bm}

\usepackage{array}
\usepackage{booktabs}
\usepackage{threeparttable}

\usepackage{siunitx}
\usepackage{physics}
\usepackage{amsmath}
\usepackage{tikz}
\usepackage{mathdots}
\usepackage{yhmath}
\usepackage{cancel}
\usepackage{color}
\usepackage{multirow}
\usepackage{amssymb}
\usepackage{gensymb}
\usepackage{tabularx}
\usepackage{extarrows}
\usepackage{booktabs}
\usetikzlibrary{fadings}
\usetikzlibrary{patterns}
\usetikzlibrary{shadows.blur}
\usetikzlibrary{shapes}

\usepackage{color}
\definecolor{LinkColor}{rgb}{0.75,0.0,0.2}

\usepackage{hyperref}
\hypersetup{
	pdfauthor={good guys},
	pdftitle={good title},
	colorlinks=true,
	citecolor=LinkColor,
	linkcolor=LinkColor,
	urlcolor=LinkColor,
}

\usepackage{listings}
\definecolor{lightgray}{gray}{1}

\begin{document}
\title{Symmetry restoration and quantum Mpemba effect in many-body localization systems}
\author{Shuo Liu}
\thanks{These two authors contributed equally to this work.}
\affiliation{Institute for Advanced Study, Tsinghua University, Beijing 100084, China}
\affiliation{Department of Physics, Princeton University, Princeton, New Jersey 08544, USA}

\author{Hao-Kai Zhang}
\thanks{These two authors contributed equally to this work.}
\affiliation{Institute for Advanced Study, Tsinghua University, Beijing 100084, China}
\affiliation{Institute of Physics, Chinese Academy of Sciences, Beijing 100190, China}

\author{Shuai Yin}
\affiliation{School of Physics, Sun Yat-sen University, Guangzhou 510275, China}

\author{Shi-Xin Zhang}
\email{shixinzhang@iphy.ac.cn}
\affiliation{Institute of Physics, Chinese Academy of Sciences, Beijing 100190, China}

\author{Hong Yao}
\email{yaohong@tsinghua.edu.cn}
\affiliation{Institute for Advanced Study, Tsinghua University, Beijing 100084, China}

\date{\today}

\begin{abstract}
Non-equilibrium dynamics of quantum many-body systems has attracted increasing attention owing to a variety of intriguing phenomena absent in equilibrium physics. A prominent example is the quantum Mpemba effect, where subsystem symmetry is restored more rapidly under a symmetric quench from a more asymmetric initial state. In this work, we investigate symmetry restoration and the quantum Mpemba effect in many-body localized systems for a range of initial states. We show that symmetry can still be restored in the many-body localization regime without approaching thermal equilibrium. Moreover, we demonstrate that the quantum Mpemba effect emerges universally for any tilted product state, in contrast to chaotic systems where its occurrence depends sensitively on the choice of the initial state. We further provide a theoretical analysis of symmetry restoration and the quantum Mpemba effect using an effective model for many-body localization. Overall, this paper fills an important gap in establishing a unified understanding of symmetry restoration and the quantum Mpemba effect in generic many-body systems, and it advances our understanding of many-body localization.
\end{abstract}

\maketitle


\vspace{3mm}
\noindent\textbf{1. Introduction}\\

Non-equilibrium physics harbors various counterintuitive phenomena and has attracted increasing attention. One famous example is the Mpemba effect~\cite{EBMpemba_1969}, namely, hot water freezes faster than cold water under identical conditions. This effect has been identified and investigated in various classical systems~\cite{PhysRevLett.119.148001, doi:10.1073/pnas.1701264114, PhysRevX.9.021060, kumarExponentiallyFasterCooling2020, bechhoeferFreshUnderstandingMpemba2021, doi:10.1073/pnas.2118484119, walker2023mpembaeffecttermsmean, walker2023optimaltransportanomalousthermal, bera2023effectdynamicsanomalousthermal, PhysRevLett.131.017101, malhotra2024doublempembaeffectcooling} and open quantum systems~\cite{PhysRevB.100.125102, PhysRevA.110.022213, PhysRevLett.131.080402, PhysRevA.106.012207, PhysRevLett.127.060401, PhysRevE.108.014130, PhysRevLett.133.010403, strachan2024nonmarkovianquantummpembaeffect, zhangObservationQuantumStrong2025, PhysRevResearch.6.033330, PhysRevLett.133.140404, PhysRevResearch.3.043108, PhysRevLett.133.136302, srivastav2024familyexactinexactquantum, xu2025expeditedthermalizationdynamicsincommensurate, wei2025quantummpembaeffectdissipative}. Recently, a quantum version of the Mpemba effect in isolated systems has been proposed~\cite{aresEntanglementAsymmetryProbe2023} where subsystem U(1) symmetry starting from a more asymmetric initial state can be restored faster than that from a more symmetric initial state under the quench of a symmetric Hamiltonian. As reviewed in Refs.~\cite{aresQuantumMpembaEffects2025, yuQuantumMpembaEffects2025}, this novel phenomenon is dubbed the quantum Mpemba effect (QME) and has been extensively investigated in integrable systems~\cite{Murciano_2024, 10.21468/SciPostPhys.15.3.089, Chalas_2024, PhysRevB.109.184312, PhysRevLett.133.010401, PhysRevB.110.085126}, free dissipative systems~\cite{Caceffo_2024, PhysRevB.111.104312}, chaotic systems~\cite{liu2024symmetry, 5d6p-8d1b, PRXQuantum.6.010324, yu2025symmetrybreakingdynamicsquantum,yu2025quantumpontusmpembaeffectsreal, chang2024imaginarytime}, and quantum simulator experiments~\cite{PhysRevLett.133.010402, xu2025observationmodulationquantummpemba}. Furthermore, the QME has also been extended to the restoration of other symmetries, including the non-Abelian SU(2) symmetry~\cite{liu2024symmetry} and the translation symmetry~\cite{PhysRevB.111.L140304}. More importantly, the underlying mechanisms of QME in both integrable and chaotic systems have been established attributing to the distinct charge transport properties~\cite{PhysRevLett.133.010401} and quantum thermalization speeds~\cite{liu2024symmetry} associated with different initial states, respectively. 

On a different front, many-body localization (MBL)~\cite{annurev:/content/journals/10.1146/annurev-conmatphys-031214-014726, annurev:/content/journals/10.1146/annurev-conmatphys-031214-014701, https://doi.org/10.1002/andp.201700169, RevModPhys.91.021001, ALET2018498} is one of the most
important cornerstones for non-equilibrium physics. Although there is an ongoing debate that MBL might only be a transient phenomenon as reviewed in Ref.~\cite{sierant2024_MBLreview} (see also Refs.~\cite{PhysRevLett.118.196801, PhysRevB.100.104204, PhysRevE.102.062144, PhysRevB.102.064207, Panda_2019, PhysRevLett.124.243601, PhysRevB.103.024203, PhysRevB.105.224203, PhysRevB.105.174205, PhysRevX.13.011041}),
the phenomenology of the MBL regimes within accessible finite-size systems is well-established.
In the presence of sufficiently strong disorder~\cite{PhysRevB.77.064426, PhysRevB.82.174411} or quasiperiodic potentials~\cite{PhysRevB.87.134202}, 
the finite-size isolated interacting system violates the eigenstate thermalization hypothesis ~\cite{PhysRevA.43.2046, PhysRevE.50.888, d2016quantum, rigolThermalizationItsMechanism2008a, Deutsch_2018} and exhibits various exotic behaviors, including the logarithmic spread of entanglement~\cite{PhysRevLett.109.017202, PhysRevLett.110.260601, PhysRevB.95.024202, https://doi.org/10.1002/andp.201600318, FAN2017707, https://doi.org/10.1002/andp.201600332, PhysRevB.96.174201, Chen2025subsystem} and emergent local integrals of motion~\cite{PhysRevLett.111.127201, PhysRevB.90.174202}.
The paradigmatic MBL Hamiltonian respects the U(1) symmetry, but the interplay between symmetry restoration and MBL regime has not been studied before. A natural question that arises is whether the U(1) symmetry can be restored in the MBL regime as the system fails to thermalize. A companion further question is whether the QME exists in the MBL regime. More importantly, a theoretical understanding of the presence or absence of symmetry restoration and QME in the MBL regime is strongly needed.

\begin{table*}[t]
  \begin{threeparttable}
    \caption{Main results of symmetry restoration and QME in MBL and chaotic systems}
    \label{tab:1}
    \begin{tabular}{c c c c}
      \toprule
      \textbf{Systems and initial states} & \textbf{Symmetry restoration} & \textbf{QME} \\
      \midrule
      MBL from any tilted product states \& 
      & Finite-size symmetry  
      & Always present \\[4pt]
      chaotic system from tilted ferromagnetic states & broken crossover & \\
      \midrule
      Chaotic system from other tilted product states 
      & Restoration for any $\theta$
      & State-dependent \\[2pt]
      \bottomrule
    \end{tabular}
    \begin{tablenotes}
      \item 
    \end{tablenotes}
  \end{threeparttable}
\end{table*}

In this paper, we investigate the U(1) symmetry restoration and the associated QME starting from various tilted product states in the thermal and MBL regimes via adjusting the strength of the disorder. In the thermal regime, the QME is present and absent for the tilted ferromagnetic state and tilted N\'eel state respectively, similar to what has been observed in U(1)-symmetric random circuits~\cite{liu2024symmetry, 5d6p-8d1b}, which can be understood through the lens of quantum thermalization. In the MBL regime, upon varying accessible system sizes, there is a tendency that the symmetry can also be restored in the thermodynamic limit, which provides a nontrivial example for the long-time evolved state that restores the symmetry but without reaching thermal equilibrium. The associated symmetry restoration timescale grows exponentially with subsystem size. More importantly, the emergence of the QME is universal in the MBL regime, independent of the choice of the initial tilted product states. In contrast, in chaotic systems, distinct types of initial states can nevertheless exhibit the same late-time behavior in symmetry-restoration dynamics, although the presence of QME remains state dependent. For instance, while both the tilted Néel state and the tilted ferromagnetic state with a middle domain wall ultimately display similar dynamical features, the QME is absent in the former but present in the latter~\cite{liu2024symmetry}. These unexpected results in MBL systems indicate a distinct underlying mechanism compared to those in integrable and chaotic systems. 

To theoretically understand the mechanism behind symmetry restoration and QME in the MBL regime, we consider the corresponding effective model based on the emergent local integrals of motion~\cite{PhysRevLett.111.127201, PhysRevB.90.174202}. In the long time limit, the degrees of symmetry breaking can be analyzed analytically. The results are identical for different initial tilted product states in MBL systems and are also consistent with the results from chaotic systems starting from initial tilted ferromagnetic states. However, the results in chaotic systems with other initial states show different patterns.
Consequently, MBL quench from any tilted product states and chaotic quench from tilted ferromagnetic states share similar symmetry restoration behaviors including the presence of QME
while QME might be absent in chaotic quench from other tilted product states. 
The main results are summarized in Table~\ref{tab:1}. See the Supplementary material for details. We further perform a direct numerical simulation and observe the QME in the symmetry restoration dynamics under the quench of the MBL effective model.

\vspace{3mm}
\noindent\textbf{2. Materials and methods}\\

We consider the following one-dimensional interacting Aubry-Andr\'e (AA) model~\cite{PhysRevB.87.134202, PhysRevB.96.075146, PhysRevX.7.031061, PhysRevB.96.060203,BarLev_2017,doi:10.1073/pnas.1800589115, PhysRevLett.120.175702, PhysRevB.96.104205, PhysRevLett.121.206601,zhang2019strong, PhysRevB.107.024204, PhysRevB.110.184209}, 
\begin{eqnarray}
    \label{eq:MBL}
    H &=& -J_{\text{hopping}}\sum_{i} (\sigma^{x}_{i} \sigma^{x}_{i+1} + \sigma^{y}_{i} \sigma^{y}_{i+1} ) \\ \nonumber
    &+& V\sum_{i} \sigma^{z}_{i} \sigma^{z}_{i+1}  
    + \sum_{i} W_{i} \sigma^{z}_{i},
\end{eqnarray}
where $\sigma_{i}^{x,y, z}$ are the Pauli matrices at site $i$, $J_{\text{hopping}}$ is the hopping strength which we set to 1 as the unit of energy,
$V$ is the strength of interaction fixed to $\frac{1}{2}$ unless otherwise specified, $W_{i} = W\cos(2 \pi \alpha i + \phi) $ is the quasiperiodic potential with strength $W$, $\alpha = \frac{\sqrt{5}+1}{2}$, and $\phi$ is a random phase to be averaged. We use the open boundary conditions throughout the work. There is an MBL regime with large $W$ (see the Supplementary material for details) for accessible finite-size systems. We have also investigated symmetry restoration and QME in the interacting model with random potentials~\cite{PhysRevB.82.174411, PhysRevB.77.064426, PhysRevB.91.081103, PhysRevB.99.104205}, and the qualitative behaviors remain the same (see the Supplementary material for details).

To quantify the degrees of symmetry breaking in subsystem $A$, we employ the entanglement asymmetry (EA)~\cite{aresEntanglementAsymmetryProbe2023} which has been extensively studied as a measure of symmetry breaking in various physical contexts~\cite{fossatiEntanglementAsymmetryCFT2024, PhysRevD.109.065009, capizziEntanglementAsymmetryOrdered2023, Capizzi_2024, PhysRevD.110.L061901, Khor2024confinementkink, 10.1093/ptep/ptaf080}. This quantity is defined as \begin{eqnarray}
    \Delta S_{A}  = S(\rho_{A, Q})   - S(\rho_{A}),
\end{eqnarray}
i.e., the difference between the von Neumann entropy of the reduced density matrix of subsystem $A$ chosen as the leftmost $N_{A}$ sites, $\rho_{A}$, and that of $\rho_{A, Q} = \sum_{q} \Pi_{q} \rho_{A} \Pi_{q}$ where $\Pi_{q}$ is the projector to the charge sector with $Q_{A}=\sum_{i \in A} \sigma_{i}^{z}=q$, namely, $\rho_{A,Q}$ retains only the block-diagonal components of $\rho_A$. EA is non-negative by definition and only vanishes when $\rho_{A}$ is block diagonal for the subsystem charge sectors, i.e., $\rho_{A}$ is U(1) symmetric. Therefore, $\Delta S_A=0$ is a necessary but not sufficient condition for thermal equilibrium. In the theoretical analysis, we utilize R\'enyi-2 EA $\Delta S^{(2)}_{A}$ by replacing von Neumann entropy with R\'enyi-2 entropy for simplicity, which shares qualitatively the same behaviors as $\Delta S_A$ and is experimentally relevant~\cite{PhysRevLett.133.010402}. 

The initial states are chosen as tilted product states. Two typical initial states include tilted ferromagnetic states (TFS) and tilted N\'eel states (TNS):
\begin{eqnarray}
    \vert \psi_{0}(\theta) \rangle = 
    \left\{  
             \begin{array}{lr}  
             e^{-i \frac{\theta}{2} \sum_{j} \sigma_{j}^{y}} \vert 0 \rangle^{\otimes N}, \  \ \ \ \ \ \text{TFS}
             &  \\  
             e^{-i\frac{\theta}{2} \sum_{j} \sigma_{j}^{y}} \vert 01 \rangle^{\otimes N/2}, \ \ \text{TNS}  
             \end{array}  
\right.
\end{eqnarray}
where $N$ is the system size and $\theta$ is the tilt angle controlling the degree of the initial symmetry breaking. EA of these two types of initial states is $\Delta S_{A}=0$ when $\theta=0$ and increases monotonically with larger $\theta$ until it reaches the maximal value at $\theta=\pi/2$.

After choosing a specific initial state, the system evolves under a quench of the Hamiltonian given by Eq.~\eqref{eq:MBL}. The reduced density matrix of subsystem $A$ at time $t$ is $\rho_{A}(t) = \tr_{\bar{A}} (e^{-iHt} \vert \psi_{0}(\theta)\rangle \langle \psi_{0}(\theta) \vert e^{iHt})$ where $\bar{A}$ is the complementary subsystem to $A$. We calculate the EA dynamics of subsystem $A$ averaged over different random phases $\phi$ to investigate the symmetry restoration and the QME. 

\vspace{3mm}
\noindent\textbf{3. Results}\\

\begin{figure}[ht]
\centering
\includegraphics[width=0.48\textwidth, keepaspectratio]{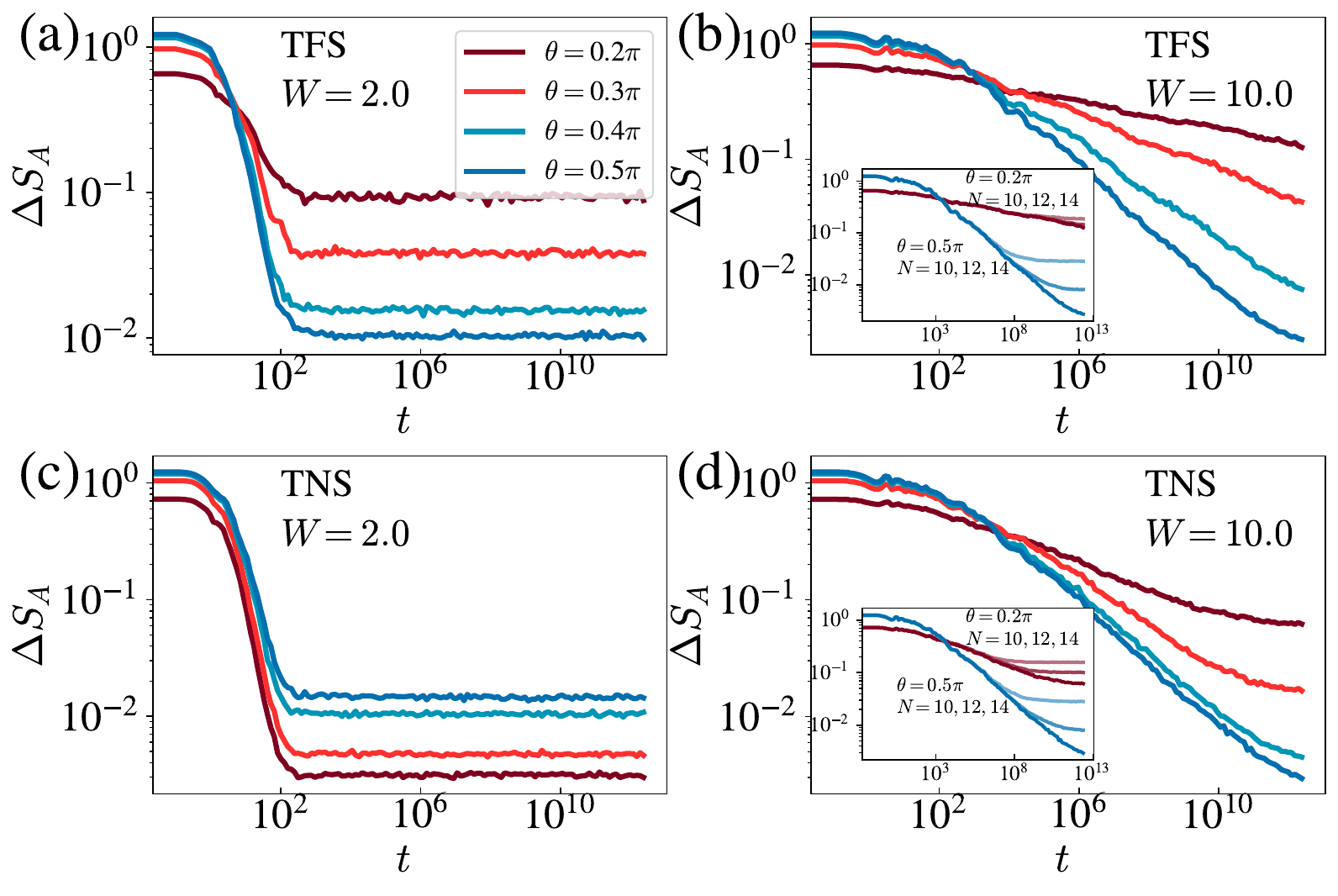}
\caption{EA dynamics averaged over different random phases $\phi$ with $N=14$ and $N_{A}=3$, i.e., subsystem $A=[1,2,3]$. The initial states of (a)(b) and (c)(d) are TFS and TNS respectively. For TFS, QME consistently appears for all values of $W$. For TNS, QME emerges exclusively in the MBL regime. The insets of panels (b) and (d) show the EA dynamics with fixed $N_{A}=3$ and varying $N$. The QME remains robust with a nearly constant timescale as the total system sizes increase.}
\label{fig:QPN12A3}
\end{figure}

We employ python packages {\sf TensorCircuit}~\cite{Zhang2023tensorcircuit} and {\sf QuSpin}~\cite{10.21468/SciPostPhys.2.1.003, 10.21468/SciPostPhys.7.2.020} to perform numerical simulations, where {\sf {T}ensor{C}ircuit} package provides a convenient method to prepare required initial states by applying quantum gates. The averaged EA dynamics with initial TFS and TNS are shown in Fig.~\ref{fig:QPN12A3}a-d. We note that the qualitative behaviors are consistent for different subsystem $A$ as long as the subsystem size $N_{A} < N/2$ (see the Supplementary material for details) and the subsystem symmetry in general cannot be restored when $N_A$ exceeds half of the total system size~\cite{PhysRevD.110.L061901}.

In the thermal regime with small $W=2.0$, as shown in Fig.~\ref{fig:QPN12A3}a and c, the QME is present for the initial TFS but absent for the initial TNS. The initial state dependence of the QME in chaotic systems has been observed in the U(1)-symmetric random circuits~\cite{liu2024symmetry, 5d6p-8d1b}, which can be understood through the lens of quantum thermalization, namely, the thermalization speed is slower in charge sectors with a smaller Hilbert-space dimension. $\rho_{A}$ of the initial TFS and TNS are both symmetric with $\theta=0$ and become more U(1) asymmetric with increasing tilt angle $\theta$. However, when $\theta=0$, TFS and TNS reduce to ferromagtic state $\vert 0 \rangle^{\otimes N} $ and N\'eel state $\vert 01 \rangle^{\otimes N/2}$ respectively. The former belongs to the smallest charge sector while the latter resides in the largest half-filling sector. Therefore, for $\rho_{A}$ of TFS, the weights of the smaller charge sectors
decrease as $\theta$ increases, i.e., the more asymmetric initial state has a faster thermalization speed~\cite{liu2024symmetry}. Consequently, the QME is anticipated. On the contrary, for $\rho_{A}$ of TNS, the weights of the smaller charge sectors increase with increasing $\theta$, i.e., the more asymmetric initial state has a slower thermalization speed~\cite{liu2024symmetry}. Therefore, the EA with a more asymmetric initial state remains larger than that with a more symmetric initial state under the quench, and thus the QME is absent.

In contrast, deep in the MBL regime with sufficiently large $W=10.0$, QME always occurs regardless of the choice of initial states as shown in Fig.~\ref{fig:QPN12A3}b and d. The symmetry restoration dynamics in the MBL regime are not only distinct from those observed in the thermal regime as discussed above but also show different late-time behaviors compared to integrable systems where symmetry cannot be restored for initial TNS~\cite{10.21468/SciPostPhys.15.3.089}. This distinction further underscores the uniqueness and significance of investigating symmetry restoration and QME in the MBL regime. Moreover, although the late-time EA saturates to a nonzero value within accessible finite-size systems, it decreases as the system size increases as shown in insets in Fig.~\ref{fig:QPN12A3}b and d,  indicating the symmetry restoration in the thermodynamic limit.

It is worth noting that the timescale of QME, defined as the EA crossing between different tilt angles $\theta$, in the MBL regime would increase exponentially with the subsystem size $N_{A}$ due to the logarithmic lightcone~\cite{PhysRevLett.109.017202, PhysRevB.95.024202, https://doi.org/10.1002/andp.201600318, FAN2017707, https://doi.org/10.1002/andp.201600332, PhysRevB.96.174201}  whereas the QME timescale in integrable and chaotic systems scales polynomially with $N_{A}$~\cite{5d6p-8d1b}. This observation is further supported by the results in the insets of Fig. \ref{fig:QPN12A3} where the QME timescale remains unchanged for the same subsystem size and different total sizes. Although the possible deviation from logarithmic lightcone at long times with increasing system size has been reported~\cite{PhysRevB.108.134204}, a much longer time is still required to observe the QME in the MBL regime in finite size systems, which explains the experimentally nonvisible QME in a disordered interacting system due to the constraint evolution time reached~\cite{PhysRevLett.133.010402}. To observe QME in the current experimental platforms, we can choose a smaller subsystem size $N_{A}$. See the Supplementary material for more discussions. 
 Moreover, it is well-known that the system in the MBL regime keeps a memory of the initial state, e.g., the non-vanishing charge imbalance starting from a N\'eel state. In other words, 
  quantities such as charge imbalance for the initial tilted N\'eel states and the associated late time states both increase with increasing $\theta$. The order-reversal behavior of EA associated with the QME studied in this work is not inconsistent with this memory characteristic of the MBL regime since EA is a non-local observable and the initial information is encoded in the diagonal part of the disorder averaged late-time subsystem density matrix (see the Supplementary material for details).

\begin{figure}[t]
\centering
\includegraphics[width=0.48\textwidth, keepaspectratio]{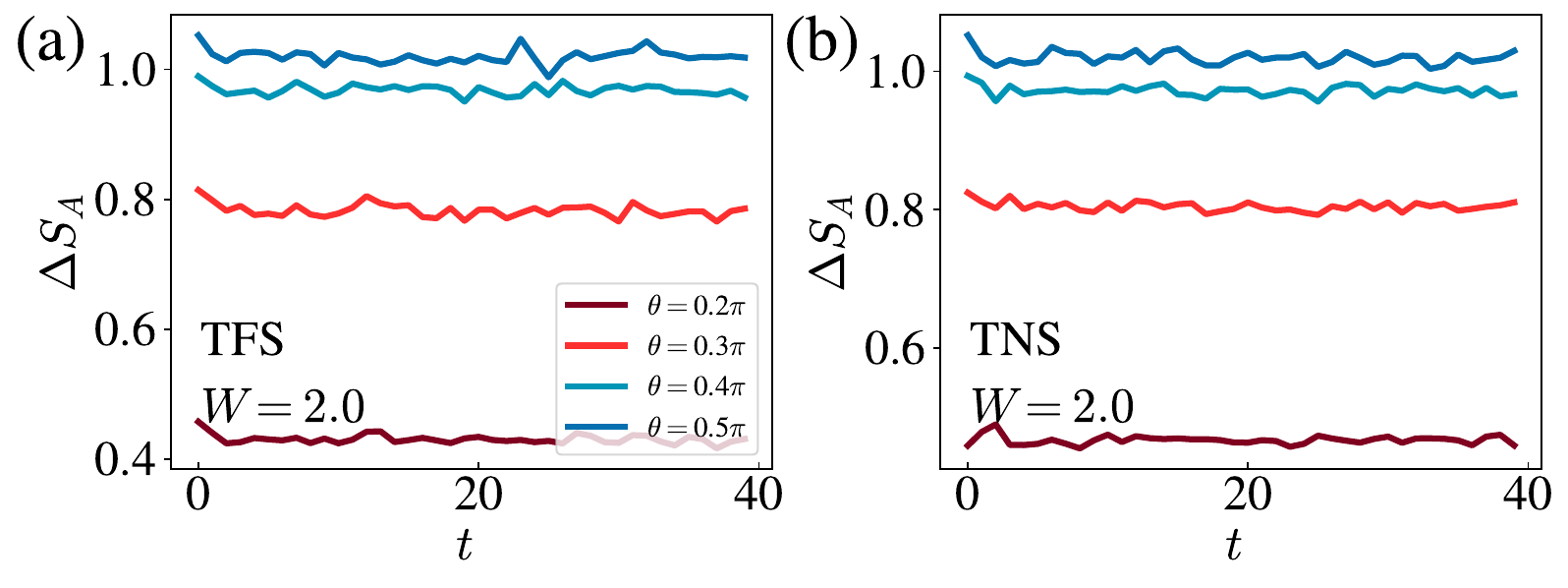}
\caption{EA dynamics in the Anderson localization phase with $V=0$ and $W=2.0$ with initial (a) TFS and (b) TNS. Here, $N=80$ and $N_{A}=10$.}
\label{fig:AndersonN80A10_main}
\end{figure}

In the absence of the interaction ($V=0$), the AA model enters the Anderson localized phase when $W>2$. As shown in Fig.~\ref{fig:AndersonN80A10_main}, the degree of symmetry breaking becomes frozen, and the EA remains essentially at its initial value. Therefore, both symmetry restoration and QME are absent in the Anderson localization phase. As detailed in the Supplementary material, these results can also be understood through the effective model of MBL discussed below by setting the coupling $J=0$.

To analytically understand the distinct behaviors of symmetry restoration in the MBL regime, we consider the fully diagonalized effective model~\cite{PhysRevB.90.174202, PhysRevLett.111.127201,imbrie2016many, PhysRevLett.117.027201, https://doi.org/10.1002/andp.201600278, https://doi.org/10.1002/andp.201600322} for the MBL regime obtained under a local unitary transformation
\begin{eqnarray}
    \label{eq:effective}
    H_{\text{eff}} = \sum_{i} h_{i} \tau_{i}^{z} + \sum_{i<j} J_{ij} \tau_{i}^{z} \tau_{j}^{z} + \cdots,
\end{eqnarray}
where $\tau_{i}^{z} = \sigma_{i}^{z} + \sum_{j,k} \sum_{\alpha, \beta = x,y,z} c^{\alpha, \beta}(i,j,k) \sigma^{\alpha}_{j}\sigma^{\beta}_{k} + \cdots$ are local integrals of motion with $c$ decaying exponentially with the distance between $i$ and $j,k$, $h_{i}$ is uniformly sampled from $[-h,h]$, and $J_{ij} = \tilde{J}_{ij}e^{-\vert i-j\vert/ \xi}$ with $\tilde{J}_{ij} \in [-J,J]$ and $\xi$ denoting the localization length. In the following analysis, we further approximate the effective model by replacing $\tau^{z}_{i}$ with $\sigma^{z}_{i}$ and neglect the higher-order terms (see more numerical results of the effective model with higher-order terms in the Supplementary material).

\begin{figure}[t]
\centering
\includegraphics[width=0.48\textwidth, keepaspectratio]{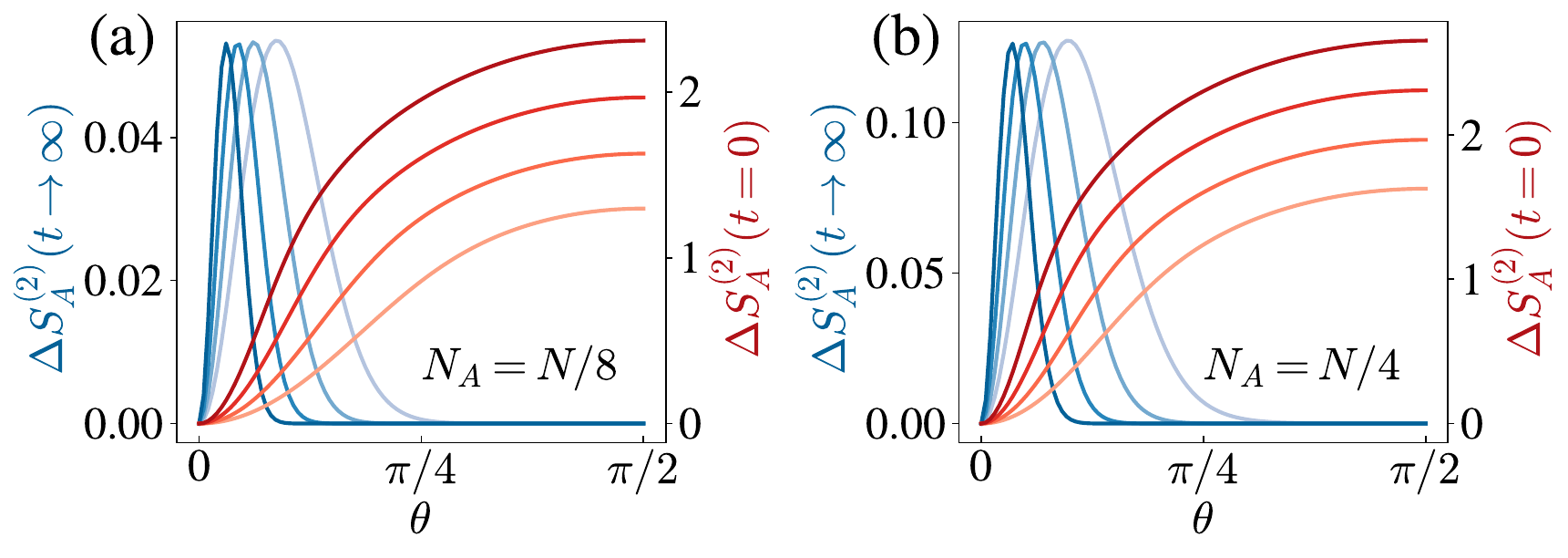}
\caption{The R\'enyi-2 EA $\Delta S_A^{(2)}$ of the initial state (red) and in the long time limit (blue). 
System sizes $N=[32,64,128,256]$ are represented with darker colors for larger $N$, and the subsystem sizes are set to $N_{A}=N/8$ and $N/4$ for panels (a) and (b), respectively.
A finite peak appears whose height remains unchanged as $N$ increases (see the Supplementary material for more discussions) indicating a finite size crossover to persistently symmetry broken phase for small $\theta\sim \frac{1}{\sqrt{N}}$.}
\label{fig:operator_spreading_main}
\end{figure}

We focus on the R\'enyi-2 EA which is experimentally relevant and 
given by
\begin{eqnarray}
    \Delta S_{A}^{(2)}(t) = \log \frac{\tr \rho_{A}^{2}(t)}{\tr \rho_{A,Q}^{2}(t)}.
\end{eqnarray}
$\rho_{A}(t)$ and $\rho_{A,Q}(t)$ can both be decomposed into a complete Pauli operator string basis $P^{\boldsymbol{\mu}} = \sigma_{0}^{\mu_0} \sigma_{1}^{\mu_1} \cdots \sigma_{N_{A}-1}^{\mu_{N_{A}-1}}$ where $\sigma_{j}^{\mu_{j}}$ is the $\mu$-type operator on $j$-th site choosing from $\{I_{j}, \sigma_{j}^{+}, \sigma_{j}^{-}, \sigma_{j}^{z} \}$, with raising and lowering operators $\sigma_{j}^{\pm} = \frac{\sigma_{j}^{x} \pm i \sigma_{j}^{y}}{\sqrt{2}}$ that breaks $U(1)$ symmetry (see the Supplementary material for details). Consequently,
\begin{eqnarray}
    \Delta S_{A}^{(2)}(t) = \log \frac{\sum_{\boldsymbol{\mu}} \vert \langle P^{\boldsymbol{\mu}} \rangle_t \vert^2}{\sum_{\boldsymbol{\mu}, [P^{\boldsymbol{\mu}}, Q_{A}] =0 }\vert \langle P^{\boldsymbol{\mu}} \rangle_t \vert^2 },
\end{eqnarray}
where the projections in $\rho_{A,Q}$ correspond to discarding the operator strings anti-commute with $Q_{A}$. Subsequently, the calculation of EA reduces to evaluating the expectation values of $P^{\mu}$. In the long time limit $t \to \infty$, we have
\begin{eqnarray}
    \label{eq:EA}
    && \Delta S^{(2)}_{A}(t \to \infty, \theta)  \\ \nonumber
    &&\approx \log (1 + (\frac{1+\cos^2(\theta)}{2})^{N-2N_{A}}
    - (\frac{1+\cos^{2}(\theta)}{2})^{N-N_{A}}),
\end{eqnarray}
which indicates that the long-time EA converges to zero in the thermodynamic limit (see Fig.~\ref{fig:operator_spreading_main}).
The late-time EA behaviors with varying tilt angle $\theta$ encode the information of both symmetry restoration and QME. On the one hand, zero late-time EA in the thermodynamic limit indicates symmetry restoration. On the other hand, the monotonic decreasing nature of late-time EA for a range of $\theta$ reflects the presence of QME in the middle times, as the order of EA with respect to $\theta$ is reversed compared to the initial monotonic increasing EA. As shown in Fig.~\ref{fig:operator_spreading_main}, when $N_{A}<N/2$ and tilt angle $\theta$ is large, i.e., the initial state is more U(1) asymmetric, the late-time R\'enyi-2 EA approaches zero and thus the symmetry can be restored in the MBL regime. However, when $\theta$ is sufficiently small, i.e., the initial state is more U(1) symmetric, the late-time R\'enyi-2 EA is finite and the height of the peak shown in Fig.~\ref{fig:operator_spreading_main}a and b stays the same with increasing system size $N$ while its position scales as $\theta \sim \frac{1}{\sqrt{N}}$, implying a finite-size crossover for persistent symmetry broken phases. Meanwhile, the late-time monotonic decreasing EA with respect to $\theta$ for $N_A<N/2$ indicates the emergence of QME.  Notably, the late-time EA results for the effective MBL model are the same for different Hamiltonian parameters and different initial product states, and also quantitatively coincide with the late-time EA results in the quantum chaotic case from TFS \cite{liu2024symmetry}.

\begin{figure}[t]
\centering
\includegraphics[width=0.48\textwidth, keepaspectratio]{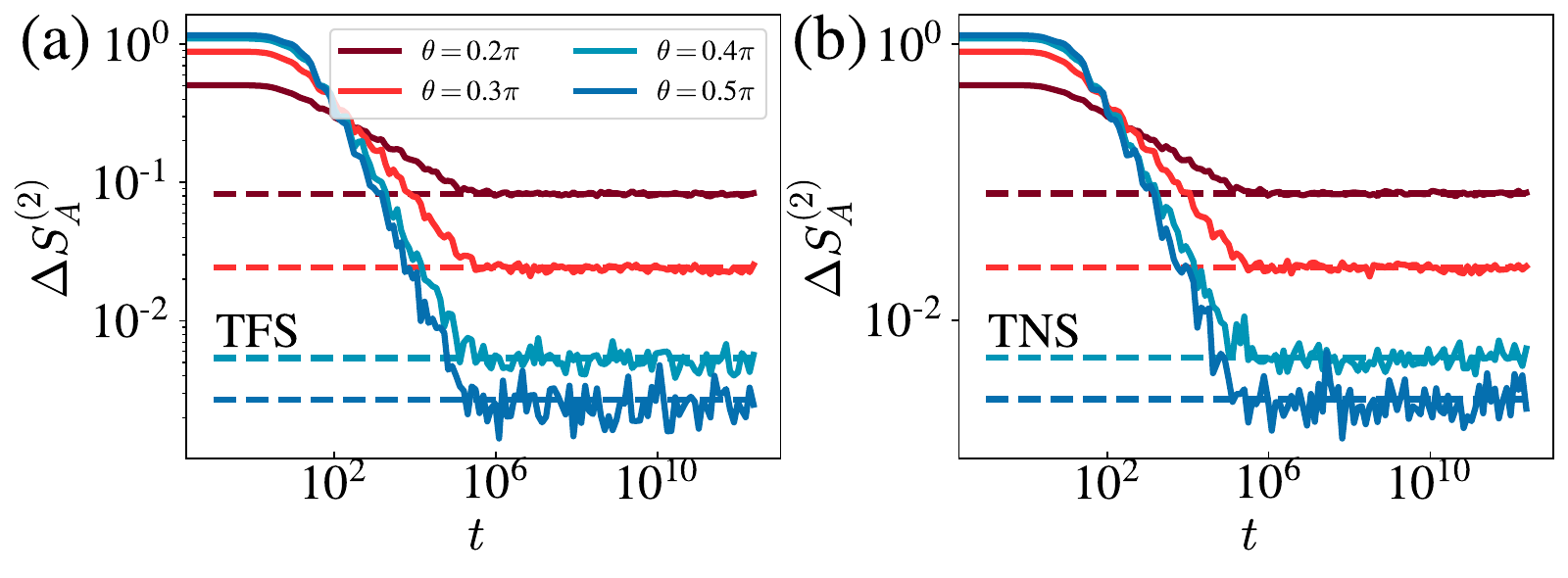}
\caption{R\'enyi-2 EA dynamics $\Delta S_A^{(2)}(t)$ of the effective model in Eq. \eqref{eq:effective} with $h=10.0$, $J=0.5$ and $\xi=1.0$. The system size is set to $N=14$ with subsystem size $N_{A}=3$. The initial states of (a) and (b) are TFS and TNS respectively. Solid lines denote numerical results, while dashed lines indicate the theoretical late-time predictions.}
\label{fig:EffectiveN12A3}
\end{figure}

Furthermore, we directly simulate the $\Delta S_{A}^{(2)}$ dynamics under the quench of the effective Hamiltonian in Eq. \eqref{eq:effective}. As shown in Fig.~\ref{fig:EffectiveN12A3}, the theoretical values in the long time limit agree well with the numerical results. More importantly, the dynamical behaviors of the effective model are qualitatively consistent with the actual dynamics, including the persistent symmetry breaking manifested as nonzero late-time EA in finite-size systems and the presence of QME regardless of the type of initial states.

Unlike quantum chaotic quench from TFS and MBL quench from generic tilted product states, late-time EA after quantum chaotic quench from other tilted product states beyond ferromagnetic states exhibits distinct behaviors -- the symmetry can always be restored without finite-size symmetry breaking peak in small $\theta$ (see the Supplementary material for details). The similarity between MBL quench from generic tilted product states with quantum chaotic quench from TFS as well as the difference between quantum chaotic quench from TFS and other tilted product states can be understood in a unified framework. The key factor is the Hilbert space dimension accessible by the initial product state. In the effective MBL model cases and ferromagnetic initial states cases with charge conservation, the accessible Hilbert space dimensions are both $O(1)$ while for other product states with a fixed ratio of charge under chaotic evolution, the accessible Hilbert space dimensions grows exponentially. Therefore, in the former case, the effective accessible Hilbert space dimension is still severely restricted with small tilt angles $\theta$, rendering persistent symmetry breaking in finite size. On the contrary, in the latter case, the exponentially large Hilbert space dimension yields sufficient relaxation to equilibrium for all $\theta$. This framework is also consistent with the mechanism for QME in chaotic systems -- small charge sectors are hard to thermalize and equilibrate only very slowly.

\vspace{3mm}
\noindent\textbf{4. Discussion and conclusion}\\

In this work, we investigate the U(1) symmetry restoration in the many-body localization regime. The U(1) symmetry can be restored in the MBL regime on an exponential time scale even though MBL forbids thermalization. Interestingly, the QME persists in the MBL regime regardless of the choice of the initial tilted product states, which is distinct from the cases in the integrable and chaotic systems. Theoretically, we derive the analytical expressions of the entanglement asymmetry for the effective MBL model in the long time limit, which are shown to be independent of the initial product states and consistent with the numerical simulation. Moreover, when the system is finite in size, the late-time subsystem symmetry cannot be fully restored and the EA remains finite when the tilt angle $\theta$ is sufficiently small but decays to zero when the tilt angle $\theta$ is large. Such late-time behaviors are reminiscent of chaotic quench from TFS, and the opposite monotonicity for EA with respect to $\theta$ between late time and early time supports the presence of the QME.

The mechanism underlying the QME in the MBL regime is fundamentally different from that in integrable and chaotic systems, characterized by different QME timescales and initial state dependence. Using the effective model, we have provided a theoretical analysis of the symmetry restoration in the MBL regime stabilized by the strong disorder. One interesting future direction is to study symmetry restoration in other types of MBL systems~~\cite{PhysRevLett.116.237203,  PhysRevB.93.094205,Sierant_2018, PhysRevLett.123.090603, PhysRevLett.122.040606, doi:10.1073/pnas.1819316116, PhysRevB.102.081115, PhysRevB.102.064206, PhysRevB.101.174204, Bhakuni_2020, PhysRevB.103.064201, PhysRevB.103.L100202, Chen2023a_z,   PhysRevLett.130.120403,PhysRevB.107.115132, Sarkar2024protectingcoherence, PhysRevLett.133.196302,PhysRevLett.119.206602, PhysRevLett.130.120403}. More importantly, this work not only offers a meaningful simulation task for current quantum devices~\cite{miTimecrystallineEigenstateOrder2022b, doi:10.1126/science.aaa7432} but also introduces a new and powerful criterion for the existence of the MBL phase: QME exists starting from TNS. Compared to the logarithmic growth of entanglement entropy, our new qualitative metric requires a much shorter evolution time and does not suffer from the subtlety induced by fitting choices.

\vspace{3mm}
\noindent\textbf{Conflict of interest}\\

The authors declare that they have no conflict of interest.

\vspace{3mm}
\noindent\textbf{Acknowledgement}\\

This work was supported in part by NSFC under Grant No. 12347107 and 12334003 (SL and HY), by
MOSTC under Grant No. 2021YFA1400100 (HY), and  by the New Cornerstone Science Foundation through the Xplorer Prize  (HY). HKZ is supported by the Postdoctoral Fellowship Program and China Postdoctoral Science Foundation under Grant No. BX20250169. SXZ acknowledge the support
from Innovation Program for Quantum Science and
Technology (2024ZD0301700) and the start-up grant
at IOP-CAS. SY is supported by the National Natural Science Foundation of China (Grants No. 12075324 and No. 12222515) and the Science and Technology Projects in Guangdong Province (Grants No. 2021QN02X561). SL's work at Princeton University was supported by the Gordon and Betty Moore Foundation
through Grant No. GBMF8685 toward the Princeton theory
program, the Gordon and Betty Moore Foundation’s EPiQS
Initiative (Grant No. GBMF11070), the Office of Naval
Research (ONR Grant No. N00014-20-1-2303), the Global
Collaborative Network Grant at Princeton University,
the Simons Investigator Grant No. 404513, the BSF
Israel US foundation No. 2018226, the NSF-MERSEC (Grant No. MERSEC DMR 2011750), the Simons
Collaboration on New Frontiers in Superconductivity,
and the Schmidt Foundation at the Princeton University.

\vspace{3mm}
\noindent\textbf{Author contributions}\\

Hong Yao and Shi-Xin Zhang conceived and designed the project. Shuo Liu, Hao-Kai Zhang and Shi-Xin Zhang performed numerical simulations and contributed theoretical analysis. All authors contributed to discussions and production of the manuscript.



\clearpage
\newpage
\widetext

\begin{center}
\textbf{\large Supplemental Material for ``Symmetry restoration and quantum Mpemba effect in many-body localization systems''}
\end{center}

\renewcommand{\thefigure}{S\arabic{figure}}
\setcounter{figure}{0}
\renewcommand{\theequation}{S\arabic{equation}}
\setcounter{equation}{0}
\renewcommand{\thesection}{\Roman{section}}

\setcounter{secnumdepth}{2} 
\setcounter{tocdepth}{1}    
\setcounter{section}{0}
\tableofcontents
\section{Numerical results for many-body localization with random potentials}
In addition to the many-body localization (MBL) model with a quasiperiodic potential, we have also investigated the symmetry restoration and the quantum Mpemba effect (QME) for the MBL model with a random potential~\cite{PhysRevB.82.174411, PhysRevB.77.064426, PhysRevB.91.081103}. Its Hamiltonian is given by
\begin{eqnarray}
    H = &-&\sum_{i} (\sigma^{x}_{i} \sigma^{x}_{i+1} + \sigma^{y}_{i} \sigma^{y}_{i+1} ) + V\sum_{i} \sigma^{z}_{i} \sigma^{z}_{i+1} + \sum_{i} W_{i} \sigma^{z}_{i},
\end{eqnarray}
where random potential $W_{i} \in [-W,W]$ is drawn from a uniform distribution and interaction strength $V$ is fixed to $1/2$. The critical strength of random potential for the many-body localization regime is $W_{c} \approx 4.3$ determined by the crossing of level spacing ratios with different system sizes, see more details in Sec.~\ref{sec:lsr}.

\subsection{Entanglement asymmetry dynamics}
\begin{figure}[ht]
\centering
\includegraphics[width=0.6\textwidth, keepaspectratio]{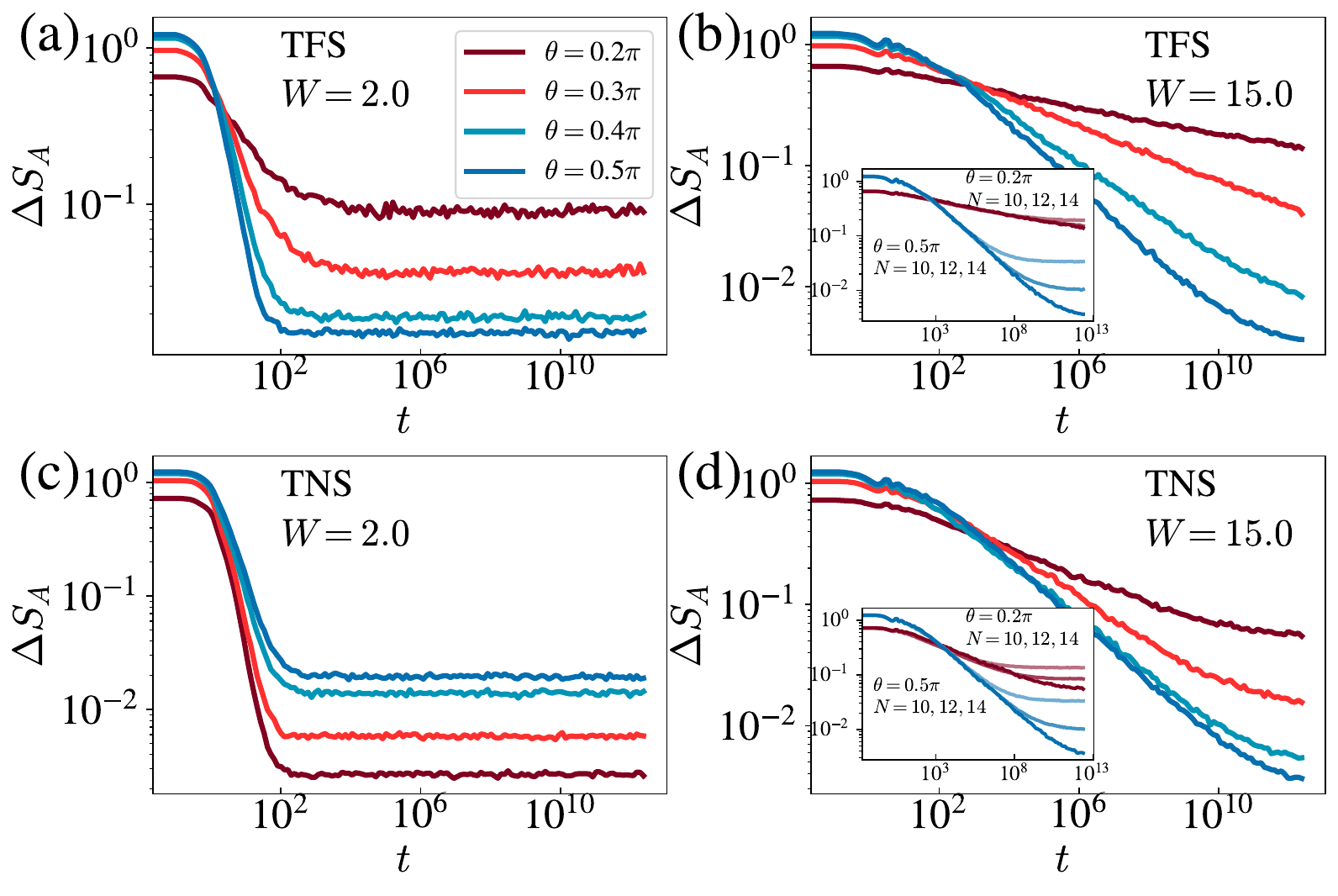}
\caption{Entanglement asymmetry dynamics with random potential averaged over different disorder realizations. We set $N=14$ and $N_{A}=3$. The initial state of (a) and (b) is TFS and the initial state of (c) and (d) is TNS. For TFS, the QME always occurs regardless of the choice of $W$, whereas for TNS, it appears only in the MBL regime. The insets of (b) and (d) show the EA dynamics with fixed $N_{A}=3$ and varying $N$.}
\label{fig:DisorderN12A3}
\end{figure}

The entanglement asymmetry (EA) dynamics with different initial tilted product states and disorder strengths $W$ are shown in Fig.~\ref{fig:DisorderN12A3}. Similar to the case of the MBL model with a quasiperiodic potential discussed in the main text, QMEs are present and absent with initial tilted ferromagnetic state (TFS) and tilted N\'eel state (TNS) respectively in the thermal regime. In contrast, in the MBL regime, QMEs persist regardless of the choice of the initial state.

\subsection{Charge imbalance dynamics}
\begin{figure}[ht]
\centering
\includegraphics[width=0.4\textwidth, keepaspectratio]{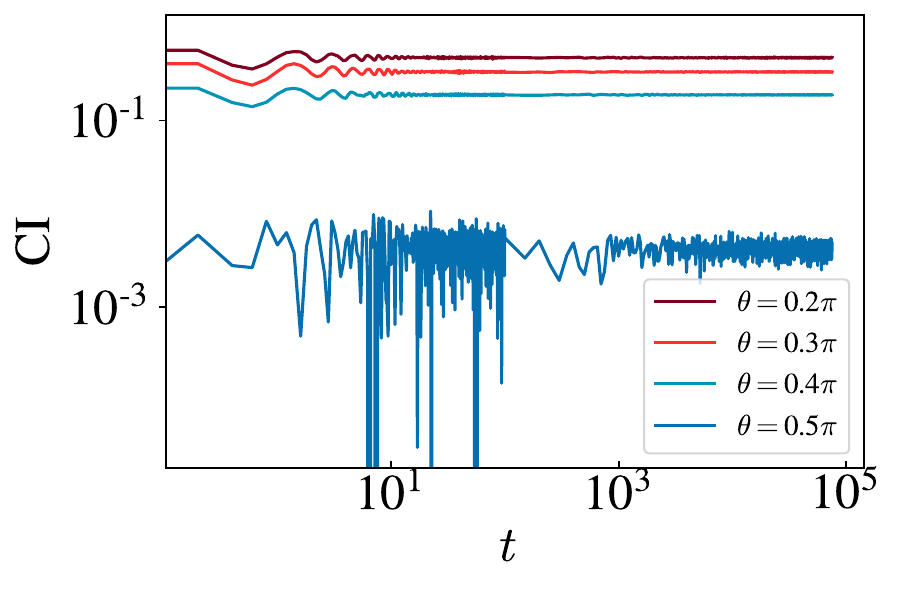}
\caption{Charge imbalance dynamics with random potential averaged over $100$ disorder realizations. Here $N=12$ and $W=20.0>W_{c}$.}
\label{fig:CI}
\end{figure}

The local information of the initial state remains in the system even after a long time evolution in the MBL regime due to the memory effects while the local observables become featureless in the thermal regime. For example, the charge imbalance of the initial N\'eel state~\cite{doi:10.1126/science.aaa7432, PhysRevLett.116.140401,PhysRevLett.119.260401,PhysRevLett.122.170403}
\begin{eqnarray}
    \text{CI}(t) = \frac{1}{N} \sum_{i=0}^{N-1} (-1)^{i} \langle \sigma^{z}_{i} \rangle_{t},
\end{eqnarray}
can be utilized to detect the MBL regime. For the TNS investigated in this work with finite initial charge imbalance, we also calculate the charge imbalance dynamics. As shown in Fig.~\ref{fig:CI} with $N=12$ and $W=20.0$, the charge imbalance remains finite at late times. Furthermore, its ordering with respect to the tilt angle $\theta$ follows that of the initial states. These properties of local observables are distinct from non-local probes such as EA, which reverses the order with respect to $\theta$ during the quench and finally approaches zero, erasing any memory of their initial states.

\section{Numerical results with smaller subsystem size for interacting  Aubry-Andr\'e model}
\begin{figure}[ht]
\centering
\includegraphics[width=0.6\textwidth, keepaspectratio]{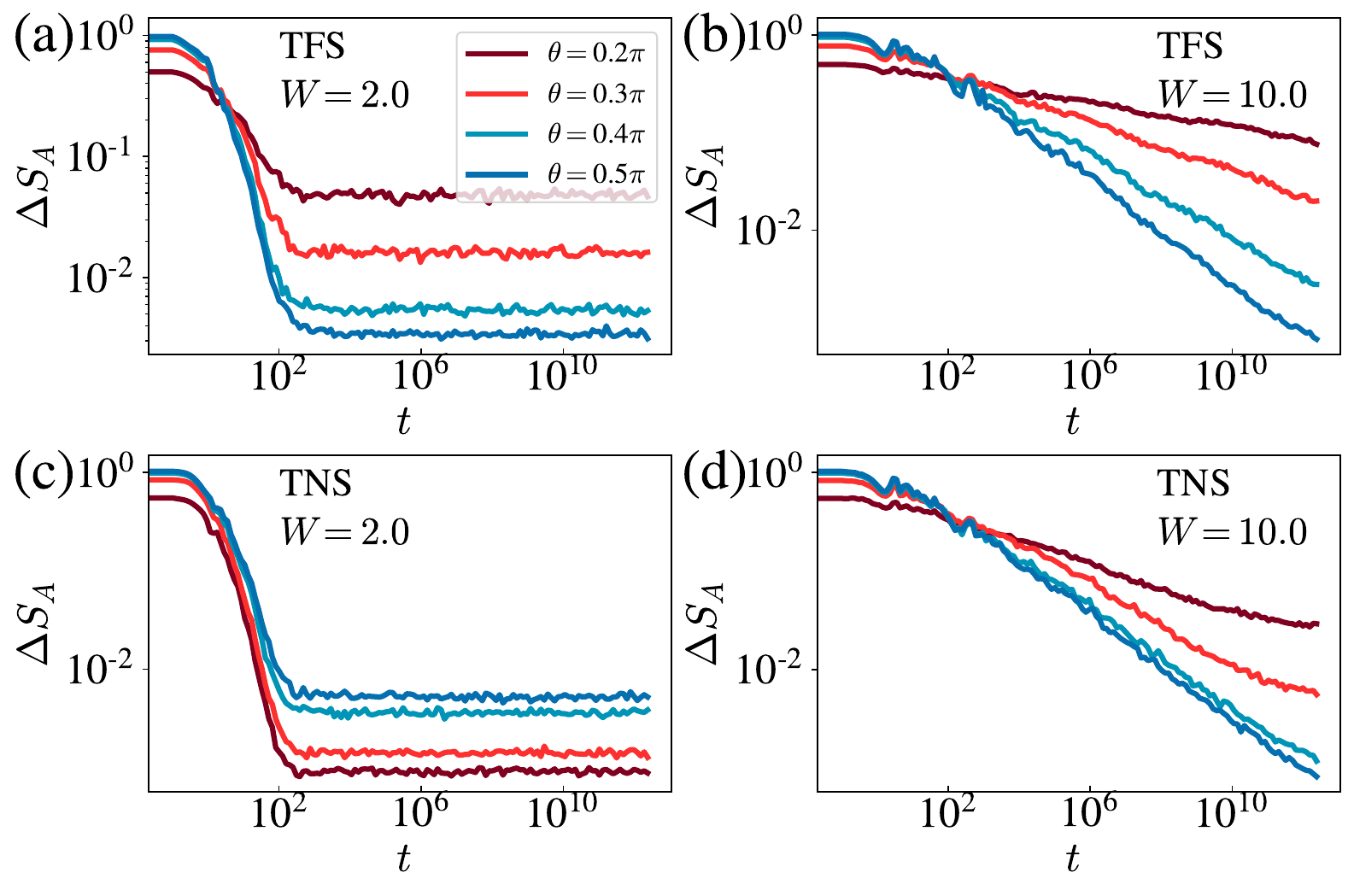}
\caption{EA dynamics averaged over different random phases $\phi$ with $N=14$ and $N_{A}=2$, i.e., subsystem $A=[1,2]$. The initial states of (a)(b) and (c)(d) are TFS and TNS respectively. The symmetry restoration dynamics is qualitatively the same as that with $N_{A}=3$.}
\label{fig:QPN12A2}
\end{figure}

In the main text, we have demonstrated the symmetry restoration dynamics and the quantum Mpemba effect under the quench of interacting AA model, where we set subsystem size $N_{A}=3$. The Mpemba time is large owing to the logarithmic spread of entanglement.
A natural question is whether the quantum Mpemba effect reported in this work can be observed in the current experimental platforms. As noted in a recent review of MBL~\cite{sierant2024_MBLreview}, $t_{\text{max}} \approx 400 (1/J_\text{hopping})$ has been reached in the experimental realization of 1D disordered Bose-Hubbard model on superconducting platforms. Therefore, the Mpemba time with $N_{A}=3$ exceeds the experimentally accessible window. Nevertheless, reducing the subsystem size provides a practical route to shorten the Mpemba time while keeping the qualitative symmetry restoration dynamics unchanged. The numerical results of entanglement asymmetry dynamics with subsystem size $N_{A} = 2$ are shown in Fig.~\ref{fig:QPN12A2}. The Mpemba time is approximately  $100(1/J_{\text{hopping}})$ that is well within the reach of the current experimental platforms.

\section{Numerical results for power-law decaying XY interacting model with random disorder}
\begin{figure}[ht]
\centering
\includegraphics[width=0.65\textwidth, keepaspectratio]{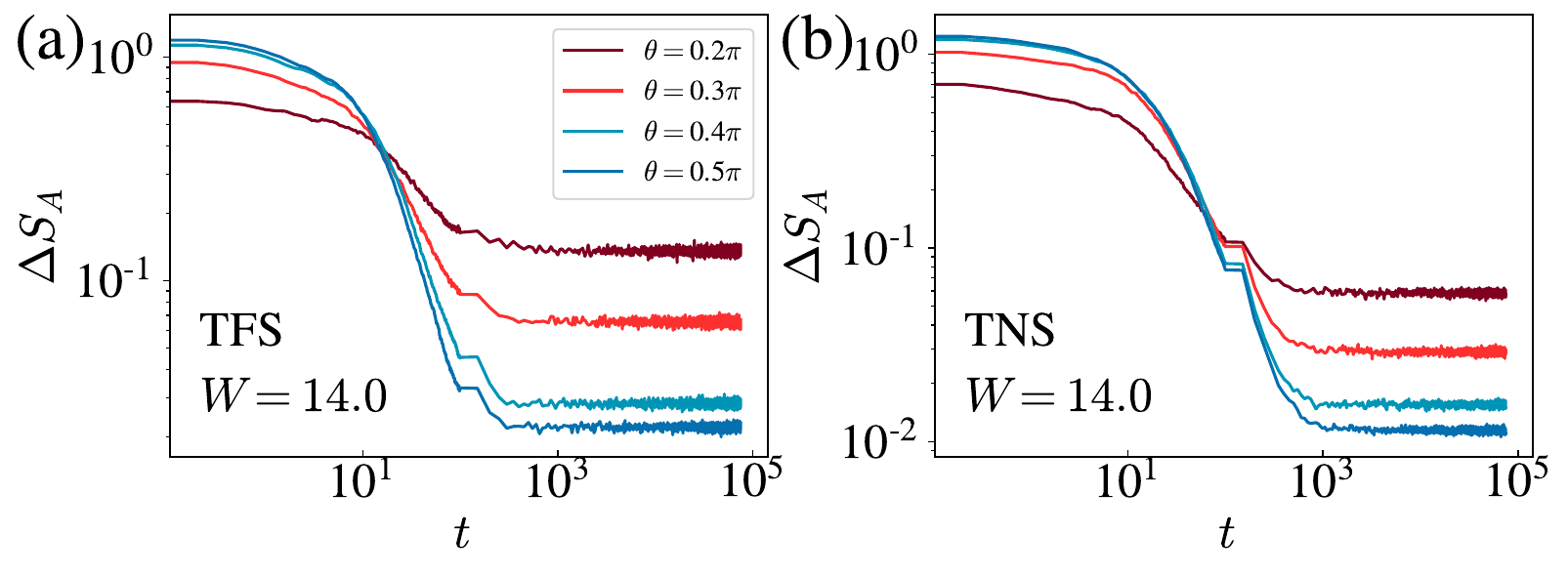}
\caption{EA dynamics under evolution with Hamiltonian shown in Eq.~\eqref{eq:longxy}. The initial state is (a) TFS and (b) TNS.
Here $N=12$, $N_{A}=3$ and $W=14.0$.}
\label{fig:longxy}
\end{figure}

The symmetry restoration for the power-law decaying XY interacting model in the presence of random disorder has recently been experimentally investigated~\cite{PhysRevLett.133.010402}. The Hamiltonian reads
\begin{eqnarray}
    \label{eq:longxy}
    H = \sum_{i>j} \frac{1}{2\vert i -j \vert^{\alpha}} (\sigma_{i}^{x} \sigma_{j}^{x} + \sigma_{i}^{y} \sigma_{j}^{y}) + \sum_{i} h_{i} \sigma_{i}^{z},
\end{eqnarray}
where $\alpha=1.0$ and $h_{i}$ uniformly distributed in $[0,W]$. The crossing of EA, i.e., the QME, was not observed within the experimental time window in Ref. \cite{PhysRevLett.133.010402} when the disorder is sufficiently strong. We have also performed additional numerical simulations of the EA dynamics under the quench of the Hamiltonian shown in Eq.~\eqref{eq:longxy} with strong disorder and open boundary conditions. Similar to the model discussed in the main text, the QME always occurs in the MBL regime with large $W$, as shown in Fig.~\ref{fig:longxy}. Therefore, the non-visible QME reported in Ref.~\cite{PhysRevLett.133.010402} is likely due to the limited time window. The exponential timescale for the presence of QME in the MBL regime may strictly limit the experimental investigation.

\section{Determining many-body localization regime via level spacing ratio} \label{sec:lsr}
To locate the critical $W_{c}$ for the many-body localization regime,
we have calculated the level spacing ratio $r$~\cite{PhysRevB.75.155111} in the half-filling charge sector. The $n$-th level spacing ratio is defined as
\begin{eqnarray}
    r_{n} = \frac{\text{min}(\Delta_{n}, \Delta_{n+1})}{\text{max}(\Delta_{n}, \Delta_{n+1})},
\end{eqnarray}
where $\Delta_{n}=E_{n+1}-E_{n}$ and $E_{n}$ is the $n$-th eigenenergy in ascending order. The level spacing ratio $r$ is obtained by averaging $r_{n}$ over energy levels and disorder realizations.

\begin{figure}[ht]
\centering
\includegraphics[width=0.4\textwidth, keepaspectratio]{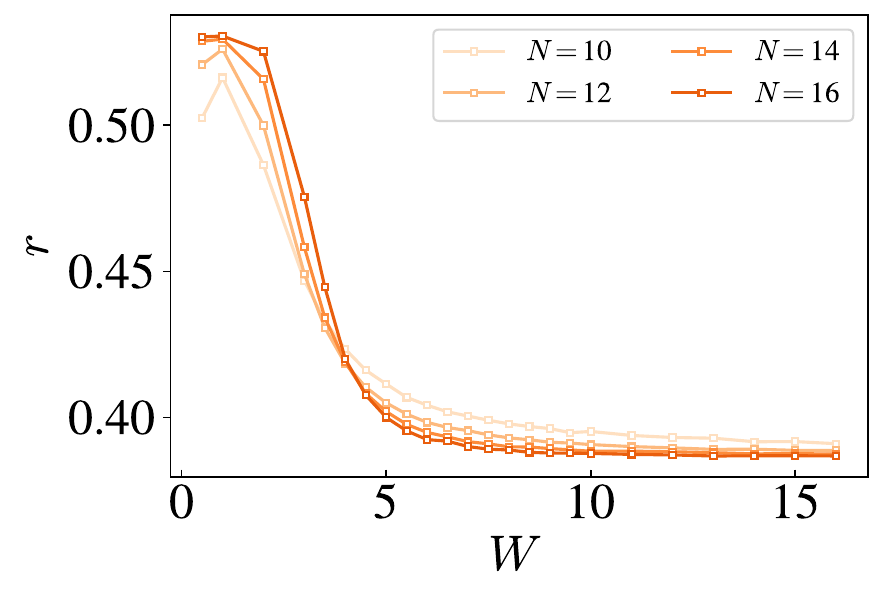}
\caption{Level spacing ratio in the half-filling sector with random potential. Each data point is averaged over $2000-10000$ disorder realizations. The critical point is $W_{c} \approx 4.3$ determined by the crossing of the level spacing ratios with $N=14$ and $N=16$.}
\label{fig:disorder_lsr}
\end{figure}

\begin{figure}[ht]
\centering
\includegraphics[width=0.4\textwidth, keepaspectratio]{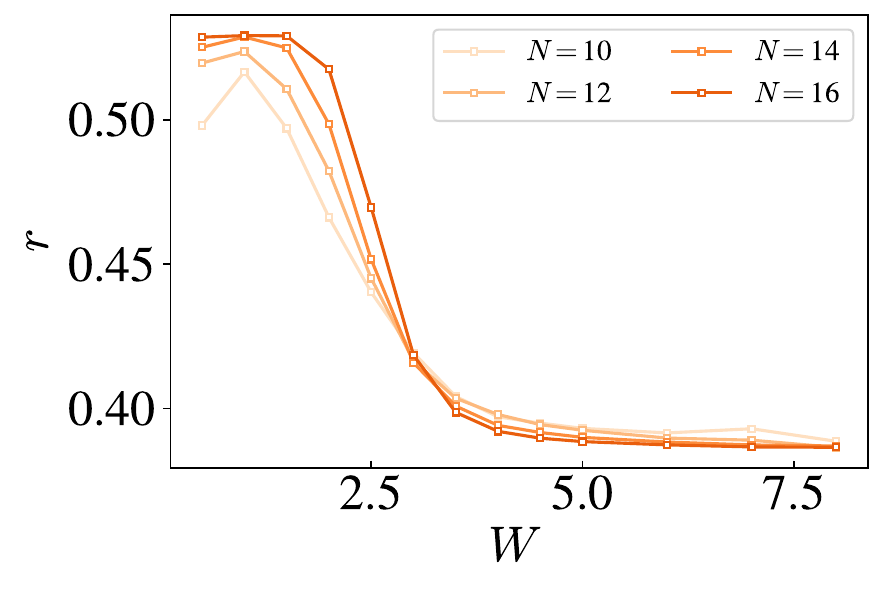}
\caption{Level spacing ratio in the half-filling sector with quasiperiodic potential. Each data point is averaged over $2000-10000$ disorder realizations. The critical point is $W_{c} \approx 3.3$ determined by the crossing of the level spacing ratios with $N=14$ and $N=16$.}
\label{fig:qp_lsr}
\end{figure}

The numerical results of the level spacing ratio with a random potential are shown in Fig.~\ref{fig:disorder_lsr} and the critical strength is $W_{c} \approx 4.3$. The numerical results of the level spacing ratio with a quasiperiodic potential are shown in Fig.~\ref{fig:qp_lsr} and the critical strength is $W_{c} \approx 3.3$.

\section{Effective model of many-body localization}
To render the analytical analysis of the quantum Mpemba effect in the many-body localization regime trackable, we consider the effective Hamiltonian~\cite{PhysRevB.90.174202, PhysRevLett.111.127201,imbrie2016many, PhysRevLett.117.027201, https://doi.org/10.1002/andp.201600278, https://doi.org/10.1002/andp.201600322} obtained under a local transformation
\begin{eqnarray}
    \label{eq:heffsm}
    H_{\text{eff}} = \sum_{i} h_{i} \tau_{i}^{z} + \sum_{ij} J_{ij} \tau_{i}^{z} \tau_{j}^{z},
\end{eqnarray}
where $\tau_{i}^{z} = \sigma_{i}^{z} + \sum_{j,k} \sum_{\alpha, \beta = x,y,z} c^{\alpha, \beta}(i,j,k) \sigma^{\alpha}_{j}\sigma^{\beta}_{k} + \cdots$ are local integrals of motion with $c$ decaying exponentially with the distance between $i$ and $j,k$, $h_{i} \in [-h,h]$, and $J_{ij} =\Tilde{J}_{ij} e^{- \vert i - j \vert / \xi }$ with $\Tilde{J}_{ij} \in [-J, J]$ and $\xi$ being the localization length. For simplicity, we further approximate the effective MBL model by replacing $\tau_{i}^{z}$ with $\sigma_{i}^{z}$. In the following sections, we present the analytical results of symmetry restoration in the Anderson phase and many-body localization regime based on this effective model.

\section{Analytical results of $\Delta S_{A}$ in the Anderson localization phase}
Before discussing the symmetry restoration and the quantum Mpemba effect in the MBL regime in the following section with the help of the effective model shown in Eq.~\eqref{eq:heffsm}, we first consider the entanglement asymmetry dynamics with $J=0$, corresponding to the Anderson localization. We choose the tilted ferromagnetic state as the initial state. The evolved state at time $t$
\begin{eqnarray}
   \vert \psi_{\theta}(t) \rangle &=& e^{-iH_{\text{eff}}t} (\cos(\theta/2) \vert 0 \rangle + \sin(\theta/2) \vert 1 \rangle)^{\otimes N} \\ \nonumber
   &=& \prod_{j=0}^{N-1} (e^{-2ih_{j}t} \cos(\theta/2) \vert 0 \rangle + \sin(\theta/2) \vert 1 \rangle)_{j},
\end{eqnarray}
remains a product state, where we have discarded an irrelevant global phase factor. Consequently, $S_{A}=0$ and thus $\Delta S_{A}=S_{A,Q}$. The reduced density matrix of subsystem $A$ of size $N_{A}$ is 
\begin{eqnarray}
    \rho_{A}(t) &=& \Tr_{\bar{A}}(\vert \psi_{\theta}(t) \rangle \langle \psi_{\theta}(t) \vert)  \\ \nonumber
    &=& \left( \prod_{j=0}^{N_{A}-1}( e^{-2  i h_{j}t} \cos(\theta/2) \vert 0 \rangle + \sin(\theta/2) \vert 1 \rangle)_{j} \right) \left( \prod_{j^{\prime}=0}^{N_{A}-1} ( e^{+2i h_{j^{\prime}}t} \cos(\theta/2) \langle 0 \vert + \sin(\theta/2) \langle 1 \vert)_{j^{\prime}} \right),
\end{eqnarray}
and thus
\begin{eqnarray}
    \rho_{A,Q} &=& \sum_{q} (\Pi_{q} \rho_{A} \Pi_{q}) \\ \nonumber 
    & =& \sum_{q} \cos^{2q}(\theta/2)\sin^{2(N_{A}-q)}(\theta/2) P_{q} O_{q} P^{\dagger}_{q},
\end{eqnarray}
where $q$ represents the charge in subsystem $A$,
$O_{q} = O^{\prime}_{q} \oplus I_{q},$ with $O^{\prime}_{q}$ being a $C_{N_{A}}^{q} \times C_{N_{A}}^{q}$ matrix whose entries are all 1 and $I_{q}$ being a $(2^{N_{A}} -C_{N_{A}}^{q}) \times (2^{N_{A}} - C_{N_{A}}^{q})$ identity matrix. We note that the order of the computational basis has been changed for simplicity. $P_{q}$ is the phase matrix of the $q$-th charge sector,
\[
P_q = \mathrm{diag}\!\Big(
e^{-2i \sum_{j \in q_{1}} h_j t},\,
e^{-2i \sum_{j \in q_{2}} h_j t},\,
\ldots,\,
e^{-2i \sum_{j \in q_{C_{N_A}^q}} h_j t},\,
\underbrace{0,0,\ldots,0}_{2^{N_A}-C_{N_A}^q}
\Big),
\]
where $q_{i}$ is the set of sites with bit $0$ in $i$-th bitstring with charge $q$ in subsystem $A$. Consequently, 
\begin{eqnarray}
    \rho_{A,Q}(t) &=& (\sum_{q} P_{q}) \rho_{A,Q}(0) (\sum_{q}  P_{q})^{\dagger}  \\ \nonumber
    &=& P \rho_{A,Q}(0) P^{\dagger},
\end{eqnarray}
where $P$ is a unitary matrix. Therefore, $S_{A,Q}(t)=S_{A,Q}(0)$ and the entanglement asymmetry remains unchanged in the Anderson localization phase, consistent with the numerical results shown in Fig.~\textcolor{LinkColor}{4} in the main text.

\section{Analytical results in the many-body localization regime}
\subsection{Connection between R\'enyi-2 entanglement asymmetry dynamics and operator spreading}

To understand the symmetry restoration and quantum Mpemba effect in many-body localized systems analytically, we can utilize the operator spreading dynamics to quantify the entanglement asymmetry~\cite{5d6p-8d1b}. Following Refs.~\cite{PhysRevX.8.021013, PhysRevX.8.031057, 5d6p-8d1b}, we introduce a suitable basis for local Hilbert space:
\begin{eqnarray}
    \sigma_{i}^{\mu} = \{I_{i}, \sigma^{+}_{i}, \sigma^{-}_{i}, \sigma_{i}^{z} \},
\end{eqnarray}
where $\sigma_{i}^{\pm} = \frac{\sigma_{i}^{x}\pm i \sigma_{i}^{y}}{\sqrt{2}}$ and $\sigma^{x,y,z}_{i}$ are Pauli operators on $i$-th qubit. These operators are traceless and orthogonal under the Frobenius inner product
\begin{eqnarray}
    \frac{1}{2} \tr [(\sigma^{\mu}_{i})^{\dagger} \sigma^{\nu}_{j}] = \delta_{ij}\delta^{\mu \nu}.
\end{eqnarray}
The operator string in subsystem $A$ of size $N_{A}$ is denoted as $P^{\boldsymbol{\mu}} = \sigma^{\mu_{0}}_{0} \sigma^{\mu_{1}}_{1} \dots \sigma^{\mu_{N_{A}-1}}_{N_{A}-1}$ and $\boldsymbol{\mu} = (\mu_{0}, \mu_{1}, \dots \mu_{N_{A}-1})$. The reduced density matrix $\rho_{A}$ can be decomposed to this operator string basis as
\begin{eqnarray}
    \label{eq:rhoAt}
    \rho_{A}(t) = \frac{1}{2^{N_{A}}} \sum_{\boldsymbol{\mu}} \langle (P^{\boldsymbol{\mu}})^{\dagger} \rangle_{t} P^{\boldsymbol{\mu}} = \frac{1}{2^{N_{A}}} (I+ \sum_{\boldsymbol{\mu}, \boldsymbol{\mu} \neq I} \langle (P^{\boldsymbol{\mu}})^{\dagger} \rangle_{t} P^{\boldsymbol{\mu}}).
\end{eqnarray}
And thus the purity of $\rho_{A}(t)$ is
\begin{eqnarray}
    \tr \rho^{2}_{A}(t) &=& \frac{1}{4^{N_{A}}} \tr \left( (I+ \sum_{\boldsymbol{\mu}, \boldsymbol{\mu} \neq I} \langle (P^{\boldsymbol{\mu}})^{\dagger} \rangle_{t} P^{\boldsymbol{\mu}}) (I+ \sum_{\boldsymbol{\nu}, \boldsymbol{\nu} \neq I} \langle (P^{\boldsymbol{\nu}})^{\dagger} \rangle_{t} P^{\boldsymbol{\nu}})  \right) \\ \nonumber
    &=& \frac{1}{2^{N_{A}}} \left( 1 + \sum_{\boldsymbol{\mu}, \boldsymbol{\mu} \neq I } |\langle P^{\boldsymbol{\mu}} \rangle_{t} |^{2}  \right) \\ \nonumber
    &=& \frac{1}{2^{N_{A}}} \sum_{\boldsymbol{\mu} } |\langle P^{\boldsymbol{\mu}} \rangle_{t} |^{2}.
\end{eqnarray}
For $\rho_{A,Q}(t)$, the operators which anti commute with $Q_{A}$ will be discarded because of $[\rho_{A,Q}(t), Q_{A}]=0$ and thus
\begin{eqnarray}
    \rho_{A,Q}(t) = \frac{1}{2^{N_{A}}} (I + \sum_{\boldsymbol{\mu}, \boldsymbol{\mu} \neq I, [P^{\boldsymbol{\mu}}, Q_{A}]=0} \langle (P^{\boldsymbol{\mu}})^{\dagger} \rangle_{t} P^{\boldsymbol{\mu}}).    
\end{eqnarray}
Therefore, the purity of $\rho_{A,Q}(t)$ is 
\begin{eqnarray}
    \tr \rho^{2}_{A,Q}(t) = \frac{1}{2^{N_{A}}} (1 + \sum_{\boldsymbol{\mu}, \boldsymbol{\mu} \neq I, [P^{\boldsymbol{\mu}}, Q_{A}]=0} |\langle P^{\boldsymbol{\mu}} \rangle_{t}|^{2}) = \frac{1}{2^{N_{A}}}  \sum_{\boldsymbol{\mu}, [P^{\boldsymbol{\mu}}, Q_{A}]=0} |\langle P^{\boldsymbol{\mu}} \rangle_{t}|^{2}.
\end{eqnarray}
Consequently, the R\'enyi-2 entanglement asymmetry is
\begin{eqnarray}
    \Delta S_{A}^{(2)} &=& S_{2}(\rho_{A,Q}) - S_{2}(\rho_{A}) \\ \nonumber
    &=& -\log (\tr \rho_{A,Q}^{2}) + \log ( \tr\rho_{A}^{2}) \\ \nonumber
    &=& \log \frac{\tr \rho_{A}^{2}}{\tr \rho_{A,Q}^{2}} \\ \nonumber
    &=& \log \frac{\sum_{\boldsymbol{\mu} } |\langle P^{\boldsymbol{\mu}} \rangle_{t} |^{2}}{ \sum_{\boldsymbol{\mu}, [P^{\boldsymbol{\mu}}, Q_{A}]=0} |\langle P^{\boldsymbol{\mu}} \rangle_{t}|^{2}} \\ \nonumber
    &=& \log \frac{f(\theta,t)}{f^{c}(\theta,t)},
\end{eqnarray}
where 
\begin{eqnarray}
    f(\theta,t) &=& \sum_{\boldsymbol{\mu} } |\langle P^{\boldsymbol{\mu}} \rangle_{t} |^{2}, \\ \nonumber
    f^{c}(\theta, t ) &=& \sum_{\boldsymbol{\mu}, [P^{\boldsymbol{\mu}}, Q_{A}]=0} |\langle P^{\boldsymbol{\mu}} \rangle_{t}|^{2},
\end{eqnarray}
and $\frac{1}{2^{N_{A}}} f(\theta,t)$ is the purity.

Having established the connection between the R\'enyi-2 entanglement asymmetry and operator spreading, we then analytically evaluate the R\'enyi-2 entanglement asymmetry of different initial states with $t=0$ and the corresponding final steady states in the long time limit $t\rightarrow \infty$.

\subsection{Tilted ferromagnetic state}
We first consider the case with the tilted ferromagnetic state as the initial state 
\begin{eqnarray}
\vert \psi_{\theta} \rangle = \left( \prod_{j=0}^{N-1}e^{-i\frac{\theta}{2} \sigma^{y}_j} \right)\vert 0\rangle^{\otimes N} = (\cos(\theta/2) \vert 0 \rangle + \sin(\theta/2) \vert 1 \rangle)^{\otimes N}.
\end{eqnarray}
The operator string with $k$ $\sigma^{+}$ operators, $l$ $\sigma^{-}$ operators and $m$ $\sigma^{z}$ is denoted as $P_{k,l,m}$. For simplicity, we assume 
\begin{eqnarray}
    P_{k,l, m} = \sigma^{+}_{0} \dots \sigma^{+}_{k-1} \sigma^{-}_{k} \dots \sigma^{-}_{k+l-1} \sigma^{z}_{k+l} \dots \sigma^{z}_{k+l+m-1} I_{k+l+m} \dots I_{N_{A}-1}.
\end{eqnarray}
We choose the computational basis and 
\begin{eqnarray}
    b &=& 1_{0} \dots 1_{k-1} 0_{k} \dots 0_{k+l-1} \{0, 1\}^{N-k-l}, \\ \nonumber
    \bar{b} &=& 0_{0} \dots 0_{k-1} 1_{k} \dots 1_{k+l-1} b_{k+l} \dots b_{N-1}
\end{eqnarray}
are two bitstrings in an $N$-qubit system where bit $0$ corresponds to the presence of a charge. We use $n_{b}$ to represent the number of $1$ in the last $N-k-l$ bits of bitstring $b$. Therefore, the number of 1 in bitstrings $b$ and $\bar{b}$ are $n_{b}+k$ and $n_{b}+l$ respectively. We use $m_{b}$ to represent the number of 1 in $b_{k+l} \dots b_{k+l+m-1}$.

The expectation square of operator string $P_{k,l,m }$ is 
\begin{eqnarray}
\label{eq:pklm}
&& | \langle \psi_{\theta} | P_{k,l, m}(t) | \psi_{\theta} \rangle |^{2}  \\ \nonumber
&=& | \sum_{b} \cos^{2N-2n_{b}-k-l}(\theta/2)\sin^{2n_{b}+k+l}(\theta/2) \langle \bar{b} | e^{i H_{\text{eff}} t} P_{k,l,m} e^{-i H_{\text{eff}} t} | b \rangle |^{2} \\ \nonumber 
&=& | \sum_{b} \cos^{2N-2n_{b}-k-l}(\theta/2)\sin^{2n_{b}+k+l}(\theta/2) e^{2i\sum_{i=0}^{k-1} h_{i} t} e^{-2i \sum_{i=k}^{k+l-1} h_{i} t} \\ \nonumber
&& e^{-2i \sum_{i=0}^{k-1} \sum_{j=k+l}^{N-1} J_{ij} (2b_{j}-1)t} e^{-2i \sum_{i=k}^{k+l-1}\sum_{j=k+l}^{N-1} J_{ij}(1-2b_{j})t} \langle \bar{b} | P_{k,l, m} | b \rangle
|^{2} \\ \nonumber 
&=& | \sum_{b} \cos^{2N-2n_{b}-k-l}(\theta/2)\sin^{2n_{b}+k+l}(\theta/2) e^{-2i \sum_{i=0}^{k-1} \sum_{j=k+l}^{N-1} J_{ij} (2b_{j}-1)t} e^{-2i \sum_{i=k}^{k+l-1}\sum_{j=k+l}^{N-1} J_{ij}(1-2b_{j})t}  \langle \bar{b} | P_{k,l,m} | b \rangle 
|^{2},
\end{eqnarray}
where we have discarded the global phase factor. 

\subsubsection{$\Delta S^{(2)}_{A}$ of the initial tilted ferromagnetic state}
When $t=0$, i.e., the initial state, 
\begin{eqnarray}
    \label{eq:ferro_p}
    | \langle \psi_{\theta} | P_{k,l, m}(t) | \psi_{\theta} \rangle |^{2} &=& 
 \left( \sum_{b} \cos^{2N-2n_{b}-k-l}(\theta/2)\sin^{2n_{b}+k+l}(\theta/2)   \langle \bar{b} | P_{k,l, m} | b \rangle \right)^{2} \\ \nonumber
    &=& 2^{k+l} \left( \sum_{n_{b}=0}^{N-k-l} \cos^{2N-2n_{b}-k-l}(\theta/2) \sin^{2n_{b}+k+l}(\theta/2) (-1)^{m_{b}} \right)^{2}  \\ \nonumber
    &=& 2^{k+l} \sin^{2k+2l}(\theta/2) \cos^{2k+2l}(\theta/2)  \\ \nonumber
    && \left( \sum_{n_{b}-m_{b}=0}^{N-k-l-m} \sum_{m_{b}=0}^{m} C_{m}^{m_{b}} (\cos^{2}(\theta/2))^{N-k-l-m-(n_{b}-m_{b})+m-m_{b}}(\theta/2) (\sin^{2}(\theta/2))^{n_{b}-m_{b}+m_{b}} (-1)^{m_{b}}\right)^{2} \\ \nonumber
    &=& 2^{k+l} \sin^{2k+2l}(\theta/2) \cos^{2k+2l}(\theta/2) \left( \sum_{m_{b}=0}^{m} C_{m}^{m_{b}} (\cos^{2}(\theta/2))^{m-m_{b}} (-\sin^{2}(\theta/2))^{m_{b}} \right)^{2} \\ \nonumber
    &=& 2^{k+l} \sin^{2k+2l}(\theta/2) \cos^{2k+2l}(\theta/2) (\cos^{2}(\theta/2) - \sin^{2}(\theta/2))^{2m} \\ \nonumber
    &=& 2^{-k-l} \sin^{2k+2l}(\theta) \cos^{2m}(\theta).
\end{eqnarray}
Consequently,
\begin{eqnarray}
    f(\theta, 0) &=& \sum_{P^{\boldsymbol{\mu}}}  \vert \langle P^{\boldsymbol{\mu}} \rangle \vert^{2}  \\ \nonumber
    &=& \sum_{k+l=0}^{N_{A}} \sum_{l=0}^{k+l} \sum_{m=0}^{N_{A}-k-l} 2^{-k-l} C_{N_{A}}^{k+l} C_{k+l}^{l} C_{N_{A}-k-l}^{m} \sin^{2k+2l}(\theta) \cos^{2m}(\theta) \\ \nonumber 
    &=& \sum_{k+l=0}^{N_{A}} \sum_{l=0}^{k+l}  2^{-k-l} C_{N_{A}}^{k+l} C_{k+l}^{l} \sin^{2k+2l}(\theta) (1+ \cos^{2}(\theta))^{N_{A}-k-l} \\ \nonumber
    &=& \sum_{k+l=0}^{N_{A}} C_{N_{A}}^{k+l} \sin^{2k+2l}(\theta) (1+ \cos^{2}(\theta))^{N_{A}-k-l} \\ \nonumber
    &=& 2^{N_{A}},
\end{eqnarray}
and 
\begin{eqnarray}
    f^{c}(\theta, 0) &=&  \sum_{P^{\boldsymbol{\mu}}, [P^{\boldsymbol{\mu}}, Q_{A}]=0}  \vert \langle P^{\boldsymbol{\mu}} \rangle \vert^{2}  \\ \nonumber
    &=& \sum_{k=0}^{N_{A}/2} \sum_{m=0}^{N_{A}-2k} 2^{-2k} C_{N_{A}}^{2k} C_{2k}^{k} C_{N_{A}-2k}^{m} \sin^{4k}(\theta) \cos^{2m}(\theta) \\ \nonumber 
    &=&  \sum_{k=0}^{N_{A}/2}  2^{-2k} C_{N_{A}}^{2k} C_{2k}^{k} \sin^{4k}(\theta) (1+ \cos^{2}(\theta))^{N_{A}-2k}.
\end{eqnarray}
Therefore, the R\'enyi-2 entanglement entropy of the initial tilted ferromagnetic states is
\begin{eqnarray}
    \Delta S_{A}^{(2)}(t=0) &=& \log \frac{f(\theta, 0)}{f^{c}(\theta, 0)} \\ \nonumber
    &=& -\log \sum_{k=0}^{N_{A}/2}  2^{-2k-N_{A}} C_{N_{A}}^{2k} C_{2k}^{k} \sin^{4k}(\theta) (1+ \cos^{2}(\theta))^{N_{A}-2k} \\ \nonumber
    &=& -\log \left( 2^{-N_{A}} (1+\cos^{2}(\theta))^{N_{A}} {}_{2}\mathcal{F}_{1}(\frac{1-N_{A}}{2}, -\frac{N_{A}}{2}, 1, \frac{\sin^{4}(\theta)}{(1+\cos^{2}(\theta))^{2}}) \right). 
\end{eqnarray}
where ${}_{2}\mathcal{F}_{1}(a,b,c,z)$ is hypergeometric function.

\subsubsection{$\Delta S^{(2)}_{A}$ of the steady state}
Now, we calculate $\Delta S^{(2)}_{A}$ of the corresponding steady state of the tilted ferromagnetic state. For $P_{k,l,m}$ with $k=l=0$ and thus $[P_{k,l,m}, H_{\text{eff}}]=0$,
\begin{eqnarray}
| \langle \psi_{\theta} | P_{0,0, m}(t) | \psi_{\theta} \rangle |^{2} = | \langle \psi_{\theta} | P_{0,0, m}(0) | \psi_{\theta} \rangle |^{2} = \cos^{2m}(\theta).
\end{eqnarray}
If $[P_{k,l,m}, H_{\text{eff}}] \neq 0$, the prefactor of the $b-b^{\prime}$ off-diagonal term shown in Eq.~\eqref{eq:pklm} is 
\begin{eqnarray}
\Re  ( e^{-2i \sum_{i=0}^{k-1} \sum_{j=k+l}^{N-1} J_{ij} (2b_{j}-1)t} e^{-2i \sum_{i=k}^{k+l-1}\sum_{j=k+l}^{N-1} J_{ij} (1-2b_{j})t}  \\ \nonumber
\times  e^{2i \sum_{i=0}^{k-1} \sum_{j=k+l}^{N-1} J_{ij} (2b^{\prime}_{j}-1)t} e^{2i \sum_{i=k}^{k+l-1}\sum_{j=k+l}^{N-1} J_{ij}(1-2b^{\prime}_{j})t} ) = \cos( \phi(J_{i,j}, b, b^{\prime}) t),
\end{eqnarray}
which is zero after averaging over different disorder realizations in the long time limit. Therefore,
\begin{eqnarray}
    \label{eq:ferro_larget_P}
   | \langle \psi_{\theta} | P_{k,l, m}(t) | \psi_{\theta} \rangle |^{2}  &=& \sum_{b} \cos^{4N-4n_{b}-2k-2l}(\theta/2) \sin^{4n_{b}+2k+2l}(\theta/2) \vert \langle \bar{b} | P_{k,l, m} | b \rangle |^{2} \\ \nonumber
   &=& 2^{k+l} \sin^{2k+2l}(\theta/2) \cos^{2k+2l}(\theta/2) \sum_{n_{b}=0}^{N-k-l}  C_{N-k-l}^{n_{b}} (\cos^{4}(\theta/2))^{N-k-l-n_{b}} (\sin^{4}(\theta/2))^{n_{b}} \\ \nonumber 
&=& 2^{k+l} \sin^{2k+2l}(\theta/2) \cos^{2k+2l}(\theta/2) (\cos^{4}(\theta/2) + \sin^{4}(\theta/2))^{N-k-l}.
\end{eqnarray}
Consequently,
\begin{eqnarray}
    f(\theta, t \to \infty) &=&  \sum_{P^{\boldsymbol{\mu}}, k+l>0}  \vert \langle P^{\boldsymbol{\mu}} \rangle \vert^{2}  + \sum_{P^{\boldsymbol{\mu}}, k+l = 0}  \vert \langle P^{\boldsymbol{\mu}} \rangle \vert^{2}  \\ \nonumber
    &=& \sum_{k+l=1}^{N_{A}} \sum_{l=0}^{k+l} 2^{N_{A}-k-l} 2^{k+l} C_{N_{A}}^{k+l} C_{k+l}^{l} \sin^{2k+2l}(\theta/2) \cos^{2k+2l}(\theta/2)(\cos^{4}(\theta/2) + \sin^{4}(\theta/2))^{N-k-l}  + \sum_{m=0}^{N_{A}} C_{N_{A}}^{m}\cos^{2m}(\theta) \\ \nonumber
    &=&  2^{N_{A}} \sum_{k^{\prime}=1}^{N_{A}} \sum_{l=0}^{k^{\prime}}  C_{N_{A}}^{k^{\prime}} C_{k^{\prime}}^{l} \sin^{2k^{\prime}}(\theta/2) \cos^{2k^{\prime}}(\theta/2)(\cos^{4}(\theta/2) + \sin^{4}(\theta/2))^{N-k^{\prime}} + (1+ \cos^{2}(\theta))^{N_{A}} \\ \nonumber 
   &=& 2^{N_{A}} \sum_{k^{\prime}=1}^{N_{A}} 2^{k^{\prime}}  C_{N_{A}}^{k^{\prime}}  \sin^{2k^{\prime}}(\theta/2) \cos^{2k^{\prime}}(\theta/2)(\cos^{4}(\theta/2) + \sin^{4}(\theta/2))^{N-k^{\prime}} + (1+ \cos^{2}(\theta))^{N_{A}}\\ \nonumber
  &=& 2^{N_{A}} \sum_{k^{\prime}=0}^{N_{A}} 2^{k^{\prime}}  C_{N_{A}}^{k^{\prime}}  \sin^{2k^{\prime}}(\theta/2) \cos^{2k^{\prime}}(\theta/2)(\cos^{4}(\theta/2) + \sin^{4}(\theta/2))^{N-k^{\prime}} \\ \nonumber
  &&- 2^{N_{A}} (\cos^{4}(\theta/2) + \sin^{4}(\theta/2))^{N} + (1+ \cos^{2}(\theta))^{N_{A}} \\ \nonumber
   &=& 2^{N_{A}} ((\cos^{4}(\theta/2) + \sin^{4}(\theta/2))^{N-N_{A}} - (\cos^{4}(\theta/2) + \sin^{4}(\theta/2))^{N}) + (1+ \cos^{2}(\theta))^{N_{A}} \\ \nonumber
&=& 2^{2N_{A}-N} (1+\cos^{2}(\theta))^{N-N_{A}} - 2^{N_{A}-N} (1+\cos^{2}(\theta))^{N} + (1+ \cos^{2}(\theta))^{N_{A}} \\ \nonumber
&=&  2^{N_{A}}(\frac{1+\cos^{2}(\theta)}{2})^{N_{A}} \left( (\frac{1+\cos^{2}(\theta)}{2})^{N-2N_{A}} - (\frac{1+\cos^{2}(\theta)}{2})^{N-N_{A}} + 1 \right)  \\ \nonumber
&=&  2^{N_{A}}(\frac{1+\cos^{2}(\theta)}{2})^{N_{A}} \left( g(\theta) + 1 \right).
\end{eqnarray}
When $N_{A} =N$, $f(\theta, t \to \infty) = 2^{N}$ and thus the purity $\frac{1}{2^{N}} f(\theta, t \to \infty) $ is $1$, consistent with the fact that the whole system is still a pure state. For the $f_{c}$,
\begin{eqnarray}
f^{c}(\theta, t \to \infty) &=&  
\sum_{P^{\boldsymbol{\mu}}, [P^{\boldsymbol{\mu}}, Q_{A}] = 0, k+l >0 } \vert \langle  P^{\boldsymbol{\mu}} \rangle \vert^{2} + (1+ \cos^{2}(\theta))^{N_{A}} \\ \nonumber
&=& \sum_{2k=2}^{N_{A}} 2^{N_{A}-2k} 2^{2k} C_{N_{A}}^{2k} C_{2k}^{k} \sin^{4k}(\theta/2) \cos^{4k}(\theta/2) (\cos^{4}(\theta/2)+\sin^{4}(\theta/2))^{N-2k}  + (1+ \cos^{2}(\theta))^{N_{A}} \\ \nonumber
&=&  2^{N_{A}} \sum_{k=1}^{N_{A}/2}  C_{N_{A}}^{2k} C_{2k}^{k} \sin^{4k}(\theta/2) \cos^{4k}(\theta/2) (\cos^{4}(\theta/2)+\sin^{4}(\theta/2))^{N-2k} + (1+ \cos^{2}(\theta))^{N_{A}} \\ \nonumber
&=& 2^{N_{A}} \sum_{k=1}^{N_{A}/2}  C_{N_{A}}^{2k} C_{2k}^{k} 2^{-4k}\sin^{4k}(\theta)  (\frac{1+\cos^{2}(\theta)}{2})^{N-2k} + (1+ \cos^{2}(\theta))^{N_{A}} \\ \nonumber
&=& 2^{N_{A}} \left( \sum_{k=1}^{N_{A}/2}  C_{N_{A}}^{2k} C_{2k}^{k} 2^{-4k}\sin^{4k}(\theta)  (\frac{1+\cos^{2}(\theta)}{2})^{N-2k} + ( \frac{1+ \cos^{2}(\theta)}{2})^{N_{A}}  \right) \\ \nonumber
&=& 2^{N_{A}} ( \frac{1+ \cos^{2}(\theta)}{2})^{N_{A}} \left( \sum_{k=1}^{N_{A}/2}  C_{N_{A}}^{2k} C_{2k}^{k} 2^{-4k}\sin^{4k}(\theta)  (\frac{1+\cos^{2}(\theta)}{2})^{N-N_{A}-2k} + 1  \right) \\ \nonumber
&=& 2^{N_{A}} ( \frac{1+ \cos^{2}(\theta)}{2})^{N_{A}} \left( g^{c}(\theta) + 1  \right)
\end{eqnarray}
In the thermodynamic limit $N \to \infty$ and $N_{A}/N<1/2$,
\begin{eqnarray}
    g^{c}(\theta)
    \ll g(\theta) \ll 1.
\end{eqnarray}
The second inequality is obvious and the first inequality can be proved as follows
\begin{eqnarray}
g^{c}(\theta) & \ll & g(\theta) \\ \nonumber
\iff \sum_{k=1}^{N_{A}/2}  C_{N_{A}}^{2k} C_{2k}^{k} 2^{-4k}\sin^{4k}(\theta)  (\frac{1+\cos^{2}(\theta)}{2})^{N-N_{A}-2k}  &\ll&   (\frac{1+\cos^{2}(\theta)}{2})^{N-2N_{A}} -  (\frac{1+\cos^{2}(\theta)}{2})^{N-N_{A}} \\ \nonumber
\iff \sum_{k=1}^{N_{A}/2}  C_{N_{A}}^{2k} C_{2k}^{k} 2^{-4k}\sin^{4k}(\theta)  (\frac{1+\cos^{2}(\theta)}{2})^{-2k} &\ll&   (\frac{1+\cos^{2}(\theta)}{2})^{-N_{A}} -  1 \\ \nonumber
\iff  \sum_{k=0}^{N_{A}/2}  C_{N_{A}}^{2k} C_{2k}^{k} 2^{-4k}\sin^{4k}(\theta)  (\frac{1+\cos^{2}(\theta)}{2})^{-2k} &\ll&   (\frac{1+\cos^{2}(\theta)}{2})^{-N_{A}} \\ \nonumber 
\iff  \sum_{k=0}^{N_{A}/2}  C_{N_{A}}^{2k} C_{2k}^{k} 2^{-4k}\sin^{4k}(\theta)  (\frac{1+\cos^{2}(\theta)}{2})^{N_{A}-2k} &\ll& 1,
\end{eqnarray}
where the last inequality is satisfied because of 
\begin{eqnarray}
 && \sum_{k=0}^{N_{A}/2}  C_{N_{A}}^{2k} C_{2k}^{k} 2^{-4k}\sin^{4k}(\theta)  (\frac{1+\cos^{2}(\theta)}{2})^{N_{A}-2k} \\ \nonumber
 &\ll& \sum_{k_1+k_2=0}^{N_{A}} \sum_{k_2=0}^{k_{1}+k_2} C_{N_{A}}^{k_1+k_2} C_{k_1+k_2}^{k_1} 2^{-2(k_1+k_2)}\sin^{2(k_1+k_2)}(\theta)  (\frac{1+\cos^{2}(\theta)}{2})^{N_{A}-k_1-k_2} \\ \nonumber 
&=& \sum_{k_1+k_2=0}^{N_{A}} C_{N_{A}}^{k_1+k_2} 2^{k_1+k_2} 2^{-2(k_1+k_2)}\sin^{2(k_1+k_2)}(\theta)  (\frac{1+\cos^{2}(\theta)}{2})^{N_{A}-k_1-k_2} \\ \nonumber
&=& \sum_{k_1+k_2=0}^{N_{A}} C_{N_{A}}^{k_1+k_2}  2^{-(k_1+k_2)}\sin^{2(k_1+k_2)}(\theta)  (\frac{1+\cos^{2}(\theta)}{2})^{N_{A}-k_1-k_2}  \\ \nonumber 
&=& \sum_{k_1+k_2=0}^{N_{A}} C_{N_{A}}^{k_1+k_2}  (\sin^{2}(\theta)/2)^{(k_1+k_2)}  (\frac{1+\cos^{2}(\theta)}{2})^{N_{A}-k_1-k_2}  \\ \nonumber
&=& 1.
\end{eqnarray}

Therefore, the R\'enyi-2  entanglement asymmetry in the long time limit is
\begin{eqnarray}
    \Delta S_{A}^{(2)} &=& \log(\frac{g(\theta)+1}{g^{c}(\theta)+1}) \\  \nonumber
    & \approx & \log(g(\theta)+1) \\ \nonumber
&=& \log \left(1 + (\frac{1+ \cos^{2}(\theta)}{2})^{N-2N_{A}} - (\frac{1+ \cos^{2}(\theta)}{2})^{N-N_{A}}\right).
\end{eqnarray}
The theoretical results of the R\'enyi-2 entanglement asymmetry of initial states and steady states are shown in Fig.~\textcolor{LinkColor}{3} in the main text, indicating the existence of the quantum Mpemba effect. 

\subsection{Tilted N\'eel state}

Now, we consider the case with the tilted N\'eel state as the initial state
\begin{eqnarray}
\vert \psi_{\theta} \rangle =  \prod_{j=0}^{N-1}e^{-i\frac{\theta}{2} \sigma^{y}_j} (\sigma^{x}_{j})^{j}  \vert 0\rangle^{\otimes N} = (\cos(\theta/2) \vert 0 \rangle + \sin(\theta/2) \vert 1 \rangle )_{\text{even}}^{\otimes N/2} \otimes (-\sin(\theta/2) \vert 0 \rangle + \cos(\theta/2) \vert 1 \rangle )_{\text{odd}}^{\otimes N/2}.
\end{eqnarray}
The operator string with $k^{e}$ $\sigma^{+}$, $l^{e}$ $\sigma^{-}$, $m^{e}$ $\sigma^{z}$ on even sites and  $k^{o}$ $\sigma^{+}$, $l^{o}$ $\sigma^{-}$, $m^{o}$ $\sigma^{z}$ on odd sites is denoted as $P_{(k,l,m)^{e}, (k,l,m)^{o}}$.  For simplicity, we assume,
\begin{eqnarray}
    P_{(k,l,m)^{e}, (k,l,m)^{o}} = P_{(k,l,m)^{e}} P_{(k,l,m)^{o}},
\end{eqnarray}
where 
\begin{eqnarray}
    P_{(k,l,m)^{e}} &=& \sigma^{+}_{0} \dots \sigma^{+}_{2k^{e}-2} \sigma^{-}_{2k^{e}} \dots \sigma^{-}_{2k^{e}+2l^{e}-2} \sigma^{z}_{2k^{e}+2l^{e}} \dots \sigma^{z}_{2k^{e}+2l^{e}+2m^{e}-2} I_{2k^{e}+2l^{e}+2m^{e}} \dots I_{N-2}, \\
     P_{(k,l,m)^{o}} &=& \sigma^{+}_{1} \dots \sigma^{+}_{2k^{o}-1} \sigma^{-}_{2k^{o}+1} \dots \sigma^{-}_{2k^{o}+2l^{o}-1} \sigma^{z}_{2k^{o}+2l^{o}+1} \dots \sigma^{z}_{2k^{o}+2l^{o}+2m^{o}-1} I_{2k^{o}+2l^{o}+2m^{o}+1} \dots I_{N-1}.
\end{eqnarray}
Suppose
\begin{eqnarray}
    b = b^{e}b^{o} = \left(1_0 \dots 1_{2k^{e}-2} 0_{2k^{e}} \dots 0_{2k^{e}+2l^{e}-2} \{0, 1\}^{N/2-k^{e}-l^{e}} \right) \left(1_1 \dots 1_{2k^{o}-1} 0_{2k^{o}+1} \dots 0_{2k^{o}+2l^{o}-1} \{0, 1\}^{N/2-k^{o}-l^{o}} \right)
\end{eqnarray}
and 
\begin{eqnarray}
   \bar{b} = \bar{b}^{e} \bar{b}^{o},
\end{eqnarray}
with 
\begin{eqnarray}
    \bar{b}^{e} &=& 0_0 \dots 0_{2k^{e}-2} 1_{2k^{e}} \dots 1_{2k^{e}+2l^{e}-2} b_{2k^{e}+2l^{e}} \dots b_{N-2}, \\
    \bar{b}^{o} &=& 0_1 \dots 0_{2k^{o}-1} 1_{2k^{o}+1} \dots 1_{2k^{o}+2l^{o}-1} b_{2k^{o}+2l^{o}+1} \dots b_{N-1}.
\end{eqnarray}
We use $n_{b}^{e(o)}$ to represent the number of $1$ in the last $N/2-k^{e(o)}-l^{e(o)}$ of $b^{e(o)}$ and $m_{b}^{e(o)}$ to represent the number of $1$ in the $b_{2k^{e}+2l^{e}} \dots b_{2k^{e}+2l^{e}+2m^{e}-2}$ $(b_{2k^{o}+2l^{o}+1} \dots b_{2k^{o}+2l^{o}+2m^{o}-1})$.

The expectation square of operator string $P_{(k,l,m)^{e}, (k,l,m)^{o}}$ is
\begin{eqnarray}
    &&|\langle \psi_{\theta} |P_{(k,l,m)^e, (k,l,m)^o}(t) | \psi_{\theta}\rangle |^2  \\ \nonumber
    &=& |\sum_{b} \cos^{(\frac{N}{2}-k^e-n_b^e)+(k^o+n_b^o)+(\frac{N}{2}-l^e-n_b^e)+(l^o+n_b^o)}(\theta/2) \sin^{(k^e+n_b^e)+(\frac{N}{2}-k^o-n_b^o)+(l^e+n_b^e)+(\frac{N}{2}-l^o-n_b^o)}(\theta/2) \\ \nonumber
    && (-1)^{(\frac{N}{2}-k^o-n_b^o)+(\frac{N}{2}-l^o-n_b^o)} R(t) \langle \bar{b} |P_{(k,l,m)^e, (k,l,m)^o}| b\rangle|^{2},
\end{eqnarray}
where $R(t)$ is the phase factor.

\subsubsection{$\Delta S_{A}^{(2)}$ of the initial tilted N\'eel state}
When $t=0$,
\begin{eqnarray}
    |\langle \psi_{\theta} |P_{(k,l,m)^e, (k,l,m)^o}(t) | \psi_{\theta}\rangle |^2 &=& ( \sum_{b} \cos^{(\frac{N}{2}-k^e-n_b^e)+(k^o+n_b^o)+(\frac{N}{2}-l^e-n_b^e)+(l^o+n_b^o)}(\theta/2) \\ \nonumber
    && \sin^{(k^e+n_b^e)+(\frac{N}{2}-k^o-n_b^o)+(l^e+n_b^e)+(\frac{N}{2}-l^o-n_b^o)}(\theta/2)  \\ \nonumber
     &&(-1)^{(\frac{N}{2}-k^o-n_b^o)+(\frac{N}{2}-l^o-n_b^o)} \langle \bar{b} |P_{(k,l,m)^e, (k,l,m)^o}| b\rangle )^{2} \\ \nonumber
     &=& 2^{-k^e-l^e-k^o-l^o} \sin^{2k^e+2l^e+2k^o+2l^o}(\theta) \cos^{2m^e+2m^o}(\theta) \\ \nonumber
     & = &  2^{-k-l} \sin^{2(k+l)}(\theta) \cos^{2m}(\theta),
\end{eqnarray}
which is the same as that with the tilted ferromagnetic state as shown in Eq.~\eqref{eq:ferro_p}. Moreover, due to Vandermonde's identity, the R\'enyi-2 entanglement asymmetry obtained by summing the expectation square of operator strings is the same as that of the tilted ferromagnetic state.

\subsubsection{$\Delta S^{(2)}_{A}$ of the steady state}
For $P_{(k,l,m)^{e}, (k,l,m)^{o}}$ with $k^e=l^e=k^o=l^o=0$ and thus $[P_{(k,l,m)^e,(k,l,m)^o}, H_{\text{eff}}]=0$
\begin{eqnarray}
    | \langle \psi_{\theta} |P_{(k,l,m)^e,(k,l,m)^o}(t) | \psi_{\theta} \rangle |^{2} &=& | \langle \psi_{\theta} |P_{(k,l,m)^e,(k,l,m)^o}(0) |
    \psi_{\theta} \rangle |^{2} \\ \nonumber
    &=& 2^{-k^e-l^e-k^o-l^o} \sin^{2k^e+2l^e+2k^o+2l^o}(\theta) \cos^{2m^e+2m^o}(\theta).
\end{eqnarray}
If $[P_{(k,l,m)^e,(k,l,m)^o}, H_{\text{eff}}] \neq 0$
, in the long time limit, all off-diagonal terms vanish after averaging over different disorder realizations and thus
\begin{eqnarray}
     &&| \langle \psi_{\theta} |P_{(k,l,m)^e,(k,l,m)^o}(t) | \psi_{\theta} \rangle |^{2} \\ \nonumber
     &=& \sum_{b} \cos^{2((\frac{N}{2}-k^e-n_b^e)+(k^o+n_b^o)+(\frac{N}{2}-l^e-n_b^e)+(l^o+n_b^o))}(\theta/2) \sin^{2((k^e+n_b^e)+(\frac{N}{2}-k^o-n_b^o)+(l^e+n_b^e)+(\frac{N}{2}-l^o-n_b^o))}(\theta/2)  \langle \bar{b} |P_{(k,l,m)^e, (k,l,m)^o}| b\rangle^{2} \\ \nonumber
     &=& 2^{-k^e-l^e-k^o-l^o} \sin^{2(k^e+l^e+k^o+l^o)}(\theta) (\sin^{4}(\theta/2)+\cos^{4}(\theta/2))^{N-k^e-l^e-k^o-l^o},
\end{eqnarray}
which is the same as that of the tilted ferromagnetic state as shown in Eq.~\eqref{eq:ferro_larget_P}. Therefore, the late-time R\'enyi-2 entanglement asymmetry of the tilted N\'eel states is also the same as that of the tilted ferromagnetic state and thus the QME is anticipated, consistent with the numerical results shown in the main text. 

\subsection{General tilted product state}
Besides two typical initial states focused on in the main text, the analytical analysis above can be easily extended to the cases with other tilted product states. More importantly, the R\'enyi-2 entanglement asymmetry is the same and independent of the choice of initial tilted product states. Suppose there are $N_{0}$ 0-bits and $N_{1}$ 1-bits in the product state before tilted. In the analytical analysis above for the tilted N\'eel state, the bits on even sites are $0$ and thus $N_{0}=N_{e}=N/2$, and the bits on odd sites are $1$ and thus $N_{1}=N_{o}=N/2$. We can also define the string operator $P_{(k,l,m)^{0}, (k,l,m)^{1}}$ similar to the case of tilted N\'eel state and the analytical results of R\'enyi-2 entanglement asymmetry can be obtained by replacing $(k,l,m)^{e}$ and $(k,l,m)^{o}$ with $(k,l,m)^{0}$ and $(k,l,m)^{1}$ respectively. Consequently, the R\'enyi-2 entanglement at $t=0$ and in the long time limit are both independent of the choice of the initial states. This initial state independence is significantly different from the cases in integrable and chaotic systems.

We now focus on the late-time density matrix structure of MBL evolution, which should be symmetric but not in thermal equilibrium. As mentioned in the main text, the MBL system keeps the memory of the local observable of the initial state in the time evolution. A natural question arises as what is the disorder averaged density matrix describing the system in the long time limit? We demonstrate the disorder averaged late-time state $\rho_A(t\rightarrow\infty)$ with the initial tilted ferromagnetic state and the extension to other initial states is straightforward. As shown in Eq.~\eqref{eq:rhoAt}, the reduced density matrix $\rho_{A}$ for each given disorder configuration can be decomposed into the operator string basis. Due to the random phase factor in $\langle P_{k,l,m}^{\dagger} \rangle$ as shown in Eq.~\eqref{eq:pklm}, only the operator string $P_{k,l,m}$ with $k=l=0$ has non-zero contribution to the diagonal terms of the reduced density matrix $\rho_{A}$ after the disorder average. Because $P_{0,0,m}$ commute with the effective Hamiltonian, its expectation value, i.e., the diagonal term in $\rho_{A}(t)$, remains the same as that of the initial state. Consequently, the system with the initial tilted ferromagnetic state in the long time limit is described by  
\begin{eqnarray}
\rho_{A}(t\rightarrow \infty) = \begin{pmatrix} \cos^{2}(\theta/2) & 0 \\ 0 & \sin^{2}(\theta/2) \end{pmatrix}^{\otimes N_{A}}.
\end{eqnarray}
This result is consistent with the memory effect of local observable in the MBL regime and presents a natural setting where long-time evolved state restores the symmetry but not approaches thermal equilibrium.

\subsection{Entanglement asymmetry dynamics for effective model with higher-order terms}
\begin{figure}[t]
\centering
\includegraphics[width=0.8\textwidth, keepaspectratio]{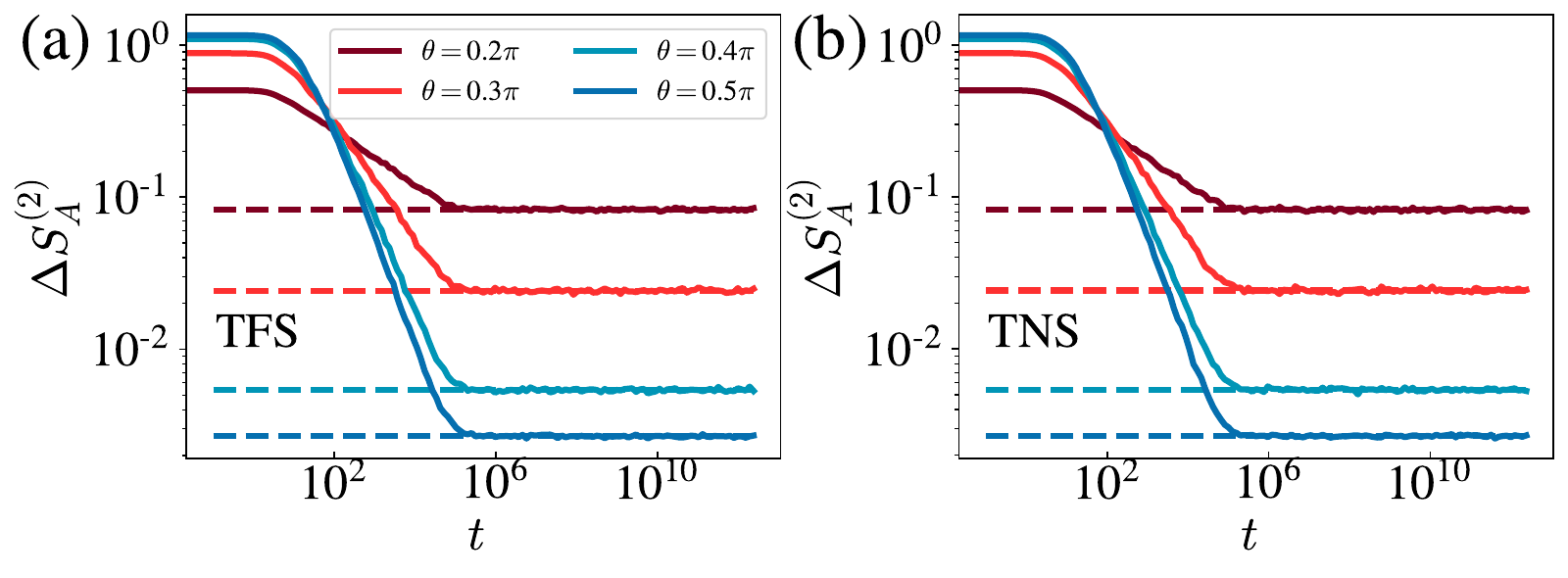}
\caption{R\'enyi-2 EA dynamics $\Delta S_A^{(2)}(t)$ of the effective model up to third-order. We choose $h=10.0$, $J=0.5$, $J^{\prime}=0.5$
and $\xi=1.0$. We set $N=14$ and $N_{A}=3$. The initial states of (a) and (b) are the TFS and the TNS respectively. The solid (dashed) line represents the numerical (theoretical late-time) results.}
\label{fig:EffectiveN14A3_3order}
\end{figure}

To verify the rationalisation of neglecting higher-order terms in the effective model, we have performed numerical simulations of the symmetry restoration dynamics of the effective model including higher-order terms, which is defined in the following:
\begin{eqnarray}
    \label{eq:effective}
    H_{\text{eff}} = \sum_{i} h_{i} \tau_{i}^{z} + \sum_{i<j} J_{ij} \tau_{i}^{z} \tau_{j}^{z} +  \sum_{i<j<k} J_{ijk} \tau_{i}^{z} \tau_{j}^{z} \tau_{k}^{z}+ \cdots,
\end{eqnarray}
where $h_{i}$ is uniformly chosen from $[-h,h]$, $J_{ij} = \tilde{J}_{ij}e^{-\vert i-j\vert/ \xi}$ with $\tilde{J}_{ij} \in [-J,J]$, $J_{ijk} = \tilde{J}_{ijk} e^{-\vert i-k\vert/\xi }$ with $\tilde{J}_{ijk} \in [-J^{\prime},J^{\prime}]$, and $\xi$ being the localization length.  As demonstrated in Fig.~\ref{fig:EffectiveN14A3_3order}, the numerical results of late-time entanglement asymmetry still agree well with the theoretical prediction and the third-order terms suppress the fluctuation in the dynamical average. Therefore, the main features of symmetry restoration have already been captured by the effective models up to second order as we considered in the main text and this approximation are valid. Theoretically, similar to the effect of the second-order terms, the higher-order terms also cause the $b-b^{\prime}$ off-diagonal terms in Eqs.~\eqref{eq:pklm} to vanish, leaving only the $b-b$ diagonal terms. The validity of the approximation is therefore confirmed.

\section{Entanglement asymmetry in random unitary circuits with different initial states}
We extend the analysis in our previous work \cite{liu2024symmetry} to give analytical late-time entanglement asymmetry results for different tilted product states beyond tilted ferromagnetic states in U(1)-symmetric random circuits, i.e. quantum chaotic systems with charge conservation.

We denote $\nu$ as the 1-doping level for a product state, i.e., the ratio of the number of qubits in $\ket{1}$ over the total qubits $N$. Thus, $\nu=0$ (or $\nu=1$) corresponds to the ferromagnetic state and $\nu=1/2$ corresponds to the N\'eel state. Suppose $N \nu$ is an integer, giving the total number of $1$ in the product states. If the initial state $\rho_0$ takes the form of the $Y$-tilted product state with $N \nu$ qubits originally in $\ket{1}$
\begin{equation}
\begin{aligned}
    e^{-\frac{i}{2}\sum_j \sigma^{y}_j\theta} \ket{0}^{\otimes N (1-\nu)}\otimes \ket{1}^{\otimes N \nu} &= \left(\cos\frac{\theta}{2}\ket{0}+\sin\frac{\theta}{2}\ket{1}\right)^{\otimes N (1-\nu)} \left(-\sin\frac{\theta}{2}\ket{0} + \cos\frac{\theta}{2}\ket{1}\right)^{\otimes N \nu} \\
    &= \sum_b \left(\cos\frac{\theta}{2}\right)^{N(1-\nu)-q_0(b)+q_1(b)} \left(\sin\frac{\theta}{2}\right)^{q_0(b)}  \left(-\sin\frac{\theta}{2}\right)^{N \nu-q_1(b)} \ket{b},
\end{aligned}
\end{equation}
where $b$ runs over all $01$-bit strings of length $N$. $q_0(b)$ and $q_1(b)$ count the number of $1$ in $b$ for the qubits originally in $\ket{0}$ and $\ket{1}$ in the untilted initial product state, respectively. Denote $\Pi_q $ as the projector onto the $q$-charge sector of the whole system. Then the weight of the initial state $\rho_0$ in the $q$-charge sector is
\begin{equation}
\begin{aligned}
    \tr(\rho_0 \Pi_q) &= \sum_b \left(\cos^2\frac{\theta}{2}\right)^{N (1-\nu) - q_0(b) + q_1(b)} \left(\sin^2\frac{\theta}{2}\right)^{q_0(b)+ N \nu - q_1(b)} \bra{b}\Pi_q\ket{b} \\
    &= \sum_b \left(\cos^2\frac{\theta}{2}\right)^{N (1-\nu) - (q_0(b)-q_1(b))} \left(\sin^2\frac{\theta}{2}\right)^{N \nu+(q_0(b)-q_1(b))} \delta_{q_0(b)+q_1(b),q} \\
    &= \sum_{q_0=0}^{q} C_{N (1-\nu)}^{q-q_0} C_{N \nu}^{q_0} \left(\cos^2\frac{\theta}{2}\right)^{N (1-\nu) - (q_0-q_1)} \left(\sin^2\frac{\theta}{2}\right)^{N \nu+(q_0-q_1)}.
\end{aligned}
\end{equation}
One can check that the number of terms is consistent by using the Chu–Vandermonde identity of binomial coefficients
\begin{equation}
    \sum_{q_0=0}^{q} C_{N (1-\nu)}^{q-q_0} C_{N \nu}^{q_0} = C_{N}^{q}.
\end{equation}
By the derivation in Ref.~\cite{liu2024symmetry}, the average purity of the reduced density matrix $\rho_A$ of the late-time evolved state on the subsystem of size $N_A$ can be expressed as
\begin{equation}\label{eq:purity_expectation_u1}
\begin{aligned}
    \mathbb{E}_\mathbb{U} [\tr(\rho_A^2)] &= \sum_{ q \neq p} \frac{\tr(\rho_0 \Pi_q ) \tr(\rho_0 \Pi_p)}{C_N^q  C_N^p} \left[ f(q,p,N_A,N_{\bar{A}}) + f(q,p,N_{\bar{A}},N_A) \right] \\
    &\quad + \sum_q  \frac{ \tr(\rho_0 \Pi_q )^2 }{C_N^q (C_N^q +1)}  \left[ f(q,q,N_A,N_{\bar{A}}) + f(q,q,N_{\bar{A}},N_A) \right],
\end{aligned}
\end{equation}
where the $f$-factor is
\begin{equation}\label{eq:f_func_def}
\begin{aligned}
    f(q,p,N_A,N_{\bar{A}})= \sum_{q'=0}^{\min\{q,p\}} C_{N_{\bar{A}}}^{q-q'} C_{N_{\bar{A}}}^{p-q'}  C_{N_A}^{q'}.
\end{aligned}
\end{equation}
Similarly, the average purity of the pruned state $\rho_{A,Q}$ is
\begin{equation}
\begin{aligned}
    \mathbb{E}_\mathbb{U} [\tr(\rho_{A,Q}^2)] &= \sum_{ q \neq p} \frac{\tr(\rho_0 \Pi_q ) \tr(\rho_0 \Pi_p)}{C_N^q  C_N^p} f(q,p,N_A,N_{\bar{A}}) \\
    &\quad + \sum_q  \frac{ \tr(\rho_0 \Pi_q )^2 }{C_N^q (C_N^q +1)}  \left[ f(q,q,N_A,N_{\bar{A}}) + f(q,q,N_{\bar{A}},N_A) \right].
\end{aligned}
\end{equation}
Based on the expressions above, one can compute the entanglement asymmetry efficiently by summing over certain products of binomial coefficients.

\begin{figure}
    \centering
    \includegraphics[width=0.65\textwidth]{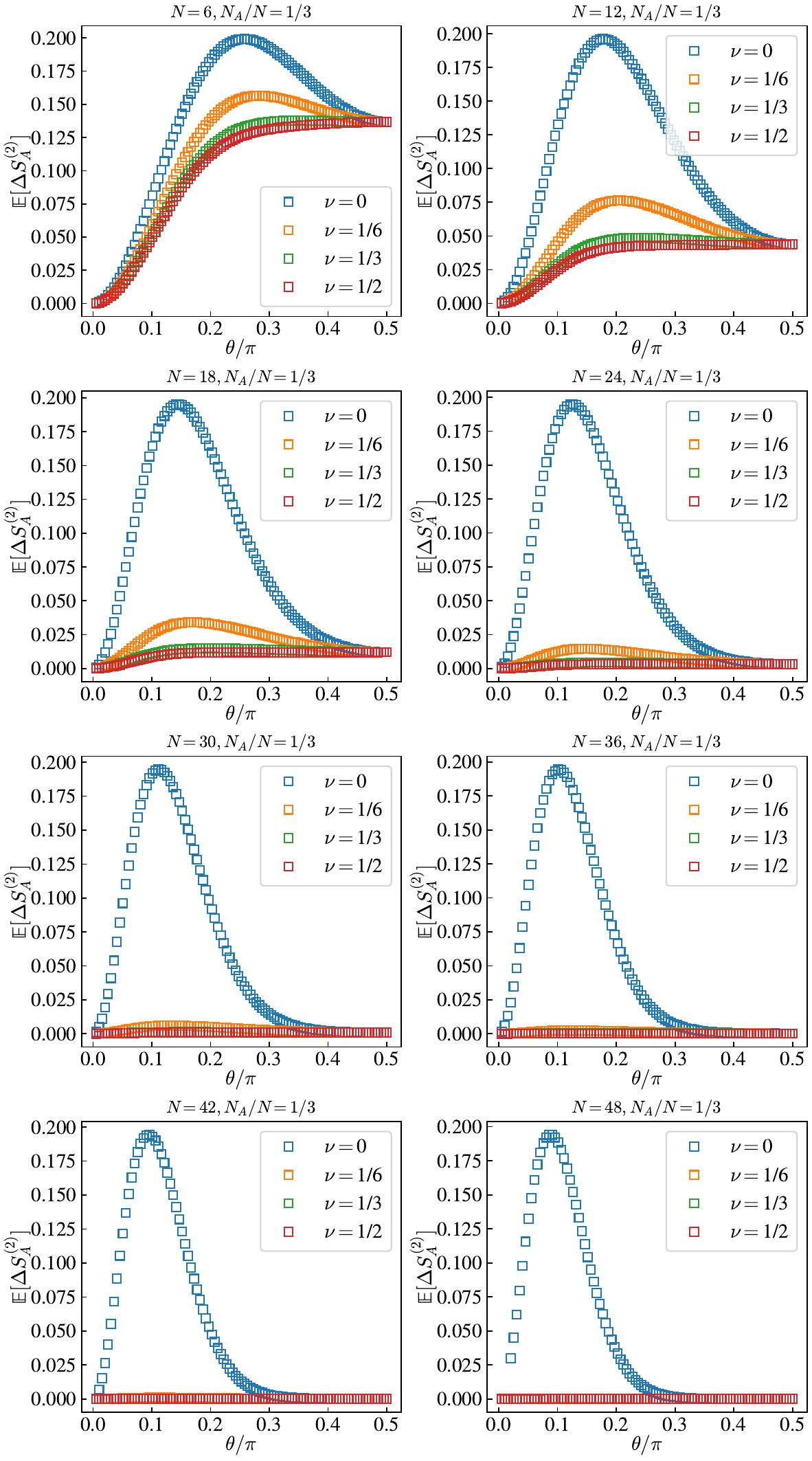}
    \caption{The average R\'enyi-2 entanglement entropy as a function of the tilt angle $\theta$ for product initial states with different $1$-doping level $\nu$ and different system sizes $N$ with $N_A/N=1/3$.}
    \label{fig:deltaS_A_doping_peak_scaling_A=1over3}
\end{figure}

As shown in Fig.~\ref{fig:deltaS_A_doping_peak_scaling_A=1over3}, we fix $N_A/N=1/3$ and compute the average R\'enyi-2 entanglement asymmetry $\mathbb{E}[\Delta S_A^{(2)}]$ for different system size $N$, tilt angle $\theta$, and $1$-doping level $\nu$. At $\nu=0$ where the initial state is a tilted ferromagnetic state, with increasing $N$, the curve of $\mathbb{E}[\Delta S_A^{(2)}]$ vs $\theta$ converges to a Gaussian-like peak with constant height and gradually leftward-shifted position of $1/\sqrt{N}$ scaling~\cite{liu2024symmetry}. That is to say, for a largely tilted ferromagnetic initial state, the symmetry is restored easily and thoroughly while for a slightly tilted one, the symmetry cannot be restored prominently for a finite-size system in the long-time limit, which can be seen as an indicator of the quantum Mpemba effect. 

\begin{figure}
    \centering
    \includegraphics[width=0.45\textwidth]{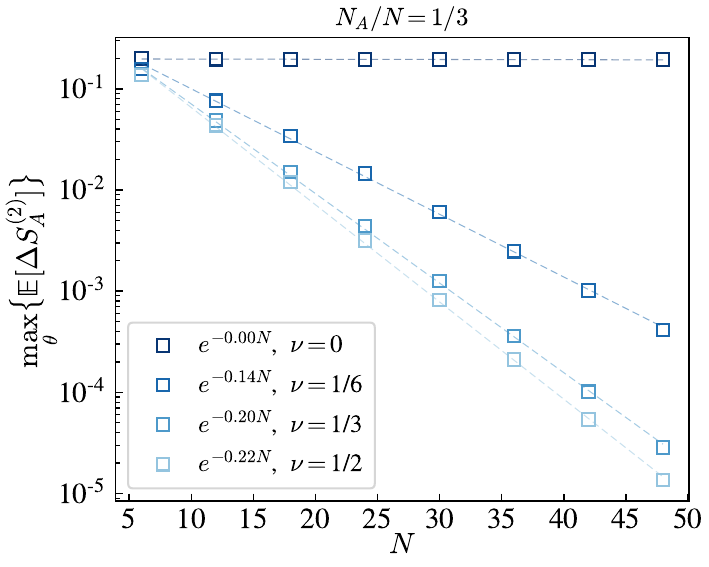}
    \caption{The maximum of the average R\'enyi-2 entanglement entropy (peak height) as a function of $N$ with $N_A/N=1/3$ for product initial states with different $1$-doping level $\nu$.}
    \label{fig:deltaS_A_doping_peak_scaling_fit_A=1over3}
\end{figure}

By contrast, at $\nu> 0$ where the initial state is still a tilted product state but with proportionable qubits in $\ket{0}$ and $\ket{1}$ ($\nu=1/2$ for the tilted N\'eel state), with increasing $N$, the overall magnitude of the curve of $\mathbb{E}[\Delta S_A^{(2)}]$ vs $\theta$ decays very fast and becomes featureless and flat at zero. More specifically, as shown by the fitting results in Fig.~\ref{fig:deltaS_A_doping_peak_scaling_fit_A=1over3}, the maximum of the curve decays exponentially with the system size $N$ for $\nu>0$ while is constant for $\nu=0$. In other words, the symmetry is restored for $\nu>0$ regardless of the values of the tilt angle $\theta$. This can be partially understood by the fact that the Hilbert space dimension for any charge sector corresponding to $\nu>0$ scales exponentially with $N$ ($C_N^{N\nu}$) while it is a constant $1$ for $\nu=0$, greatly limiting the occurrence of quantum thermalization.


\end{document}